\documentclass{aa}
\usepackage{graphicx}
\usepackage{journals}
\usepackage{natbib}
\def\HI{{H\,{\small I}}}
\def\HIt{{H\,{\scriptsize I}}}
\def\HIb{{H\,{\Large I}}}
\newcommand{\hi}{{H$\,$\footnotesize I}}
\newcommand{\shi}{\overline{\Sigma}\subhi}
\newcommand{\df}{\mbox{DEF}}

\newcommand{\subhi}{_\mathrm{H\mbox{\tiny I}}}
\newcommand{\mhi}{M\subhi}
\begin{document}
\title{The origin of \HIb-deficiency in galaxies on the outskirts of the
Virgo cluster}
\titlerunning{\HIt-deficiency on outskirts of Virgo. II}
\subtitle{II. Companions and uncertainties in distances and deficiencies}

\author{T. Sanchis \inst{1}
          \and
  G. A. Mamon \inst{2,3}          
	\and
          E. Salvador-Sol\'e\inst{1,4}
          \and
          J. M. Solanes\inst{1,4}
	  }

\offprints{T. Sanchis}

\institute{Departament d'Astronomia i Meteorologia, Universitat de Barcelona, Mart\'{\i} i
Franqu\`es 1, 08028 Barcelona, Spain\\
\email{tsanchis@am.ub.es; eduard@am.ub.es; jsolanes@am.ub.es}
\and
Institut d'Astrophysique de Paris (CNRS UMR 7095), 98 bis Bd Arago,
F-75014 Paris, France\\
\email{gam@iap.fr}
\and
GEPI (CNRS UMR 8111), Observatoire de Paris, F--92195 Meudon Cedex, France
\and
CER on Astrophysics, Particle Physics, and Cosmology, Universitat de Barcelona,  Mart\'{\i} i
Franqu\`es 1, 08028 Barcelona, Spain \\
}

\date{Received ............; accepted ...........}

\abstract{
The origin of the deficiency in neutral Hydrogen of 13
spiral galaxies lying
in the
outskirts of the Virgo cluster is reassessed.
If these galaxies have passed through the core of the cluster, their
interstellar gas should have been lost through ram pressure stripping by the
hot X-ray emitting gas of the cluster.
We analyze the positions of these \HIt-deficient and other spiral galaxies in
velocity-distance plots, in which we include our compilation of
velocity-distance data on 61
elliptical galaxies, and compare with 
simulated velocity-distance diagrams
obtained from cosmological $N$-body simulations. 
We find that $\sim20\%$ relative Tully-Fisher
distance errors are consistent with the great majority of the spirals, except
for a small number of objects, 
whose positions in the velocity-distance diagram suggest grossly incorrect
distances, which implies that the Tully-Fisher error distribution function
has non-gaussian wings. Moreover, we find that 
the distance errors may lead to an incorrect fitting of the
Tolman-Bondi solution that can generate significant errors in the
distance and especially the mass 
estimates of the cluster.
We suggest 4 possibilities for the outlying \HIt-deficient spirals (in
decreasing frequency):
 1) they  
have large relative distance errors and
are in fact close enough (at distances between 12.7 and 20.9 Mpc from us) 
to the cluster to have passed through its core and
seen their gas removed 
by ram pressure stripping; 2) their gas is converted to stars by tidal 
interactions with other galaxies; 
3) their gas is heated during recent mergers with smaller galaxies; 
and 4) they are in reality not \HI-deficient
(e.g. S0/a's misclassified as 
Sa's). 

}

\maketitle
    
\section{Introduction} \label{introduction}

During the last thirty years, substantial 
work has been carried out to quantify the
effects of the environment on the properties of galaxies, i.e. to distinguish
the characteristics of galaxies in dense, aggressive regions such as
clusters, especially cluster cores,
and 
low density zones away from clusters and groups.
The nearby Virgo cluster is an excellent candidate
for such studies, because its proximity allows
1) an easy comparison with field galaxies and 2)
the measurement of distances independent of radial velocity, providing 
three-dimensional information on cluster properties
(e.g. \citealp{YC95,Gavazzi+99,Solanes+02})
impossible to obtain in other clusters.

One particular characteristic of spiral galaxies is their neutral Hydrogen
(\HI) gas content.
It has been known for some time that a class of \emph{\HI-deficient} galaxies
exists in the 
centers of clusters, such as the Virgo cluster \citep*{CBG80},
and their deficiency, normalized to their optical
diameter and morphological type, is anti-correlated with their clustercentric
distance \citep{HG86,Solanes+01}.

\cite{CBG80} 
considered ram pressure stripping of interstellar atomic
Hydrogen by the hot diffuse intracluster gas, first proposed by 
\cite{GG72},
as one cause of the \HI\
deficiency pattern of Virgo spirals.
For galaxies moving face-on, ram pressure scales as the intracluster gas
density times the square of the 
galaxy's velocity relative to this intracluster gas, so the role of ram
pressure stripping was strengthened by the
correlation of \HI\ deficiency with the X-ray luminosity of clusters
found by \citeauthor{GH85} (\citeyear{GH85}; see however \citealp{Solanes+01}) 
and by the presence of truncated \hi\ disks observed by \cite{CvGBK90}.

It hence came as a surprise when \cite{Solanes+02} discovered spiral galaxies,
whose Hydrogen content was deficient relative to other spiral galaxies of the
same morphological type, far away
from the core of Virgo, with several
deficient spirals in the
foreground of Virgo and particularly one group of spirals 7 to 10 Mpc
(depending on the adopted distance to Virgo) behind the cluster
with one-third of its 15 spirals displaying \HI\ 
deficiencies deviating by more than $3\,\sigma$ from normalcy.

In a followup study, \cite{Sanchis+02} 
explored the possibility 
that the group of deficient spirals behind Virgo might
have plunged into the
cluster core a few Gyr ago and lost their gas by interacting with the hot
intracluster medium.
Sanchis et al.
applied a simple
dynamical model that considered not only the classical deduction of the
virgocentric infalling regime, but also the trajectories of galaxies that had
already crossed once the center of the core. This model succeeded in explaining
the overall characteristics of the observed velocity-distance diagram, 
including the envelope of the virialized 
cluster drawn by galaxies on first rebound. With the available data, they
argued, from a statistical point of view, that
one could not discard the possibility that some of the gas-deficient 
objects, including the background group in the outskirts of the
Virgo Cluster Region, were not newcomers, but have passed through the cluster
and lost their gas by
ram pressure stripping. 

We \citep{MSSS03} have recently estimated the maximum radius out to where
galaxies falling into a cluster can bounce out. We found that this maximum
rebound radius corresponds to 1 to 2.5 cluster virial radii. After estimating
the virial radius of the Virgo cluster using X-ray observations, we concluded
that the \HI-deficient galaxies located in the foreground or background of the
Virgo cluster appear to lie much 
further from the center of the cluster than these
maximum distances, and therefore that ram pressure stripping cannot account for
the \HI-deficiency of these outlying galaxies if their distances are correct.

In this paper, we investigate in further detail the origin of the
\HI-deficiency in the spiral galaxies that appear too far from the Virgo
cluster to have suffered ram pressure stripping of their interstellar gas, by
1) adding to our four-dimensional 
(angular position, distance and radial velocity) sample of spiral
galaxies, our compilation of the literature for elliptical galaxies towards
Virgo, 
leading us to identify more precisely the density
of the environment of the background deficient galaxies and
estimate more accurately 
the distance to the center of Virgo, which is a crucial ingredient for the
dynamical infall model,
and 2)
using cosmological 
$N$-body simulations that have
the advantages of their lack of observational errors, completeness, and of
defining a realistic infall pattern.

The sample of elliptical galaxies is described in Sect.~\ref{ellips}.
In Sect.~\ref{replica}, after describing the
cosmological $N$-body simulations used here,
we  
compare the simulated and
observational samples to evaluate the influence of errors in distance
estimates.
Finally in Sect.~\ref{deficiency}, we discuss possible
explanations for the \HI\ deficiency of the outlying galaxies.

\section{The inclusion of ellipticals into the Virgo galaxy sample}
\label{ellips}

Using
only late-type spirals, \cite{Sanchis+02} applied a spherical
infall model to the velocity field of the Virgo Cluster Region and obtained
two possible distances of 20 and 21 Mpc. 
Other distance estimates to the Virgo cluster, based upon spiral
galaxies, produced distances in the range 
from 16 Mpc 
\citep{TS84} to 21 Mpc \citep{ELTF00}.

An alternative way to estimate the cluster distance is by using the early-type
components of the cluster, as they are thought to be better tracers of the
center of the cluster due to morphological segregation (see e.g.
\citealp{FB94}). To do so,  we have collected as many as possible early-type
galaxies (E's, dE's and S0's) with redshift-independent  distance measurements.
Heliocentric velocities were extracted from the {\sf LEDA} catalog of the
{\sf HyperLEDA} database and
converted to 
the Local Group reference frame, while the direct distant measurements are
obtained from a variety of sources.

\begin{table}[ht]
\tabcolsep 4pt
\caption[]{Datasets contributing to elliptical sample}
\begin{center}
\begin{tabular}{lcccc}
\hline \hline
Method$^a$ & Authors$^b$ & Band & rel. accuracy & Number\\
	&	&	& (rms) &  \\
\hline
FP & G & $H$ & 0.41 & 41 \\
$L-n$& $\rm Y_{\mit L}$ & $B_{J}$ & 0.54 & ~\,9 \\
$R-n$& $\rm Y_{\mit R}$ & $B_{J}$ & 0.54 & 14 \\
SBF & T & $I$ & 0.18 & 31 \\
SBF & $\rm J_{\mit I}$ & $I$ & 0.07 & ~\,6 \\
SBF & $\rm J_{\mit K}$ & $K'$ & 0.17 & ~\,7 \\
SBF & N & $I$ & 0.15 & 15 \\
GCLF & K & $V\!-\!I$ & 0.07 & ~\,8 \\
\hline 
\end{tabular}
\end{center}
$^a$ FP = Fundamental Plane; $L-n,R-n$ = Luminosity / S\'ersic shape
and Radius / S\'ersic shape relations; SBF = Surface Brightness Fluctuations;
GCLF = Globular Cluster Luminosity Function.
$^b$ 
G = \cite{Gavazzi+99}; 
T = \cite{Tonry+01}; 
$\rm J_{\mit I}$,$\rm J_{\mit K}$ = \cite*{JTL98};
N = \cite{NT00};
K = \cite{KW01}; 
$\rm Y_{\mit L}$,$\rm Y_{\mit R}$ = \cite{YC95}.
\label{table:authors}
\end{table}

Table~\ref{table:authors} summarizes the distance 
datasets on the early-type galaxies towards Virgo.
Although
different authors use different methods
with different accuracy, 
we have not found systematic deviations
between them, and furthermore found that the spread in differences of the
distance moduli was consistent with the quoted errors, for all sets of
early-type galaxies common to two authors.
Therefore, we estimated a global distance modulus $\overline{d}$ for each
galaxy, using the maximum likelihood estimator (e.g. \citealp{Bevington92}) 
\begin{equation}
\overline{d}= \frac{\sum_i d_i  / \sigma_i^{2}} {\sum_i
1/\sigma_i^2}
\ ,
\label{likehood}
\end{equation}
where
$d_i$ are the distance moduli and $\sigma_i$ are the absolute errors on
these distance moduli.
Equation~(\ref{likehood}) supposes 
that the relative error on distance measurements is independent
of distance and that the p.d.f. of the distance moduli
(proportional to log distance) is
gaussian.
The error on the global distance modulus is (e.g. \citeauthor{Bevington92})
\begin{equation}
\sigma_{\overline{d}}^2 = {1\over \sum_i 1/\sigma_i^2} \ .
\label{sigma0}
\end{equation}
Although this estimator of the error on the distance modulus is independent
on the scatter of the measurements, we show in Appendix~\ref{appsig}
(eqs.~[\ref{xmed}] and [\ref{newsigma}]) that
equations~(\ref{likehood}) and (\ref{sigma0}) also represent the median and
the half-width (as defined by $(x_{84}-x_{16})/2$) of the likelihood
function, respectively.

Table~\ref{ellip} provides our final sample of 63 early-type galaxies.
Columns (1), (2) and (3) give the name of the galaxy, with its coordinates
in columns (4) and (5). In column (6), we give the distance (in Mpc, from eq.~[\ref{likehood}]) and its
relative error (eq.~[\ref{sigma0}])
in column (7). Column (8) gives the velocity relative to the
Local Group and column (9) provides the sources for the distance.

\begin{table*}
\label{ellip}
\caption[]{Final sample of early-type galaxies}
\begin{center}
\tabcolsep 3mm
\begin{tabular}{rcccccrcrl}
\hline
\hline
\multicolumn{3}{c}{Galaxy} & &
RA & 
Dec & 
\multicolumn{1}{c@{\hspace{2mm}}}{$D$} & 
rel. error & 
\multicolumn{1}{c@{}}{$v$} &
sources \\
\cline{1-3}
\cline{5-6}
\multicolumn{1}{c}{VCC} & 
\multicolumn{1}{c}{NGC} & 
Messier & &
\multicolumn{2}{c}{(J2000)} & 
\multicolumn{1}{c}{(Mpc)} & & 
\multicolumn{1}{c@{}}{($\rm km \,s^{-1}$)} \\
\hline
  49 & 4168 & --- & & $\rm12^h12^m17\fs3$ & $+13^\circ12'17''$ & 34.20 & 0.41 & $  2128 $ & $ \rm G $ \\ 
 220 & 4233 & --- & & $\rm12^h17^m07\fs7$ & $+07^\circ37'26''$ & 32.96 & 0.41 & $  2181 $ & $ \rm G $ \\ 
 312 & 4255 & --- & & $\rm12^h18^m56\fs1$ & $+04^\circ47'09''$ & 28.71 & 0.41 & $  1603 $ & $ \rm G $ \\ 
 319 &  --- & --- & & $\rm12^h19^m02\fs0$ & $+13^\circ58'48''$ &  7.59 & 0.38 & $  -266 $ & $ \rm Y_{\mit L}Y_{\mit R} $ \\ 
 342 & 4259 & --- & & $\rm12^h19^m22\fs2$ & $+05^\circ22'34''$ & 57.02 & 0.41 & $  2306 $ & $ \rm G $ \\ 
 345 & 4261 & --- & & $\rm12^h19^m23\fs2$ & $+05^\circ49'32''$ & 32.27 & 0.16 & $  2012 $ & $ \rm GT $ \\ 
 355 & 4262 & --- & & $\rm12^h19^m30\fs6$ & $+14^\circ52'38''$ & 14.86 & 0.41 & $  1181 $ & $ \rm G $ \\ 
 369 & 4267 & --- & & $\rm12^h19^m45\fs4$ & $+12^\circ47'53''$ & 14.66 & 0.41 & $   852 $ & $ \rm G $ \\ 
 575 & 4318 & --- & & $\rm12^h22^m43\fs3$ & $+08^\circ11'52''$ & 25.35 & 0.41 & $  1047 $ & $ \rm G $ \\ 
 648 & 4339 & --- & & $\rm12^h23^m34\fs9$ & $+06^\circ04'54''$ & 16.44 & 0.17 & $  1108 $ & $ \rm T $ \\ 
 685 & 4350 & --- & & $\rm12^h23^m57\fs7$ & $+16^\circ41'33''$ & 14.86 & 0.41 & $  1049 $ & $ \rm G $ \\ 
 731 & 4365 & --- & & $\rm12^h24^m28\fs3$ & $+07^\circ19'04''$ & 22.07 & 0.05 & $  1058 $ & $ \rm J_{\mit I}J_{\mit K}NT $ \\ 
 763 & 4374 &  84 & & $\rm12^h25^m03\fs7$ & $+12^\circ53'13''$ & 17.71 & 0.07 & $   733 $ & $ \rm GNT $ \\ 
 778 & 4377 & --- & & $\rm12^h25^m12\fs3$ & $+14^\circ45'43''$ & 15.49 & 0.41 & $  1182 $ & $ \rm G $ \\ 
 784 & 4379 & --- & & $\rm12^h25^m14\fs8$ & $+15^\circ36'26''$ & 14.19 & 0.38 & $   871 $ & $ \rm T $ \\ 
 810 &  --- & --- & & $\rm12^h25^m33\fs8$ & $+13^\circ13'30''$ & 12.36 & 0.38 & $  -516 $ & $ \rm Y_{\mit L}Y_{\mit R} $ \\ 
 828 & 4387 & --- & & $\rm12^h25^m41\fs8$ & $+12^\circ48'35''$ & 17.21 & 0.35 & $   373 $ & $ \rm GT $ \\ 
 881 & 4406 &  86 & & $\rm12^h26^m12\fs2$ & $+12^\circ56'44''$ & 17.40 & 0.04 & $  -465 $ & $ \rm J_{\mit I}J_{\mit K}KNT $ \\ 
 940 &  --- & --- & & $\rm12^h26^m47\fs1$ & $+12^\circ27'14''$ &  7.24 & 0.54 & $  1337 $ & $ \rm Y_{\mit R} $ \\ 
 953 &  --- & --- & & $\rm12^h26^m54\fs7$ & $+13^\circ33'57''$ & 18.88 & 0.38 & $  -678 $ & $ \rm Y_{\mit L}Y_{\mit R} $ \\ 
 965 &  --- & --- & & $\rm12^h27^m03\fs1$ & $+12^\circ33'39''$ &  9.29 & 0.54 & $   614 $ & $ \rm Y_{\mit R} $ \\ 
1025 & 4434 & --- & & $\rm12^h27^m36\fs7$ & $+08^\circ09'14''$ & 25.50 & 0.15 & $   897 $ & $ \rm GT $ \\ 
1030 & 4435 & --- & & $\rm12^h27^m40\fs6$ & $+13^\circ04'44''$ & 13.12 & 0.41 & $   611 $ & $ \rm G $ \\ 
1146 & 4458 & --- & & $\rm12^h28^m57\fs6$ & $+13^\circ14'29''$ & 17.97 & 0.09 & $   499 $ & $ \rm GKNT $ \\ 
1173 &  --- & --- & & $\rm12^h29^m14\fs8$ & $+12^\circ58'41''$ & 14.96 & 0.38 & $  2292 $ & $ \rm Y_{\mit L}Y_{\mit R} $ \\ 
1196 & 4468 & --- & & $\rm12^h29^m30\fs9$ & $+14^\circ02'55''$ & 15.85 & 0.13 & $   732 $ & $ \rm T $ \\ 
1226 & 4472 &  49 & & $\rm12^h29^m46\fs7$ & $+07^\circ59'59''$ & 16.27 & 0.03 & $   693 $ & $ \rm GJ_{\mit I}J_{\mit K}KNT $ \\ 
1231 & 4473 & --- & & $\rm12^h29^m48\fs9$ & $+13^\circ25'46''$ & 17.12 & 0.05 & $  2041 $ & $ \rm GKNT $ \\ 
1242 & 4474 & --- & & $\rm12^h29^m53\fs7$ & $+14^\circ04'06''$ & 15.14 & 0.41 & $  1405 $ & $ \rm G $ \\ 
1250 & 4476 & --- & & $\rm12^h29^m59\fs1$ & $+12^\circ20'55''$ & 19.08 & 0.11 & $  1767 $ & $ \rm NT $ \\ 
1253 & 4477 & --- & & $\rm12^h30^m02\fs2$ & $+13^\circ38'11''$ & 16.44 & 0.41 & $  1164 $ & $ \rm G $ \\ 
1279 & 4478 & --- & & $\rm12^h30^m17\fs5$ & $+12^\circ19'40''$ & 16.68 & 0.09 & $  1220 $ & $ \rm GNT $ \\ 
1297 &~~4486B & --- & & $\rm12^h30^m32\fs0$ & $+12^\circ29'27''$ & 16.31 & 0.03 & $   136 $ & $ \rm KN $ \\ 
1308 &  --- & --- & & $\rm12^h30^m45\fs9$ & $+11^\circ20'36''$ & 17.86 & 0.38 & $  1511 $ & $ \rm Y_{\mit L}Y_{\mit R} $ \\ 
1316 & 4486 &  87 & & $\rm12^h30^m49\fs4$ & $+12^\circ23'28''$ & 16.82 & 0.09 & $  1095 $ & $ \rm GNT $ \\ 
1321 & 4489 & --- & & $\rm12^h30^m52\fs3$ & $+16^\circ45'30''$ & 17.04 & 0.13 & $   772 $ & $ \rm J_{\mit K}T $ \\ 
1386 &  --- & --- & & $\rm12^h31^m51\fs5$ & $+12^\circ39'24''$ &  7.08 & 0.38 & $  1135 $ & $ \rm Y_{\mit L}Y_{\mit R} $ \\ 
1412 & 4503 & --- & & $\rm12^h32^m06\fs3$ & $+11^\circ10'35''$ & 11.75 & 0.41 & $  1184 $ & $ \rm G $ \\ 
1420 &  --- & --- & & $\rm12^h32^m12\fs3$ & $+12^\circ03'42''$ & 11.75 & 0.38 & $   847 $ & $ \rm Y_{\mit L}Y_{\mit R} $ \\ 
1489 &  --- & --- & & $\rm12^h33^m14\fs0$ & $+10^\circ55'43''$ & 11.30 & 0.38 & $   -94 $ & $ \rm Y_{\mit L}Y_{\mit R} $ \\ 
1535 & 4526 & --- & & $\rm12^h34^m03\fs1$ & $+07^\circ41'57''$ & 16.61 & 0.17 & $   427 $ & $ \rm GT $ \\ 
1537 & 4528 & --- & & $\rm12^h34^m06\fs2$ & $+11^\circ19'15''$ & 14.52 & 0.41 & $  1195 $ & $ \rm G $ \\ 
1539 &  --- & --- & & $\rm12^h34^m06\fs3$ & $+12^\circ44'40''$ &  8.83 & 0.54 & $  1216 $ & $ \rm Y_{\mit R} $ \\ 
1549 &  --- & --- & & $\rm12^h34^m14\fs9$ & $+11^\circ04'16''$ & 19.32 & 0.54 & $  1202 $ & $ \rm Y_{\mit R} $ \\ 
1619 & 4550 & --- & & $\rm12^h35^m30\fs7$ & $+12^\circ13'13''$ & 17.84 & 0.09 & $   208 $ & $ \rm KNT $ \\ 
1630 & 4551 & --- & & $\rm12^h35^m38\fs1$ & $+12^\circ15'49''$ & 17.24 & 0.15 & $   973 $ & $ \rm GT $ \\ 
1632 & 4552 &  89 & & $\rm12^h35^m40\fs0$ & $+12^\circ33'22''$ & 15.31 & 0.04 & $   115 $ & $ \rm GJ_{\mit I}J_{\mit K}KNT $ \\ 
1664 & 4564 & --- & & $\rm12^h36^m27\fs0$ & $+11^\circ26'18''$ & 14.86 & 0.15 & $   947 $ & $ \rm GT $ \\ 
1669 &  --- & --- & & $\rm12^h36^m30\fs6$ & $+13^\circ38'18''$ &  8.79 & 0.54 & $   427 $ & $ \rm Y_{\mit R} $ \\ 
1692 & 4570 & --- & & $\rm12^h36^m53\fs5$ & $+07^\circ14'46''$ & 13.80 & 0.41 & $  1559 $ & $ \rm G $ \\ 
1720 & 4578 & --- & & $\rm12^h37^m30\fs6$ & $+09^\circ33'16''$ & 17.94 & 0.06 & $  2102 $ & $ \rm GJ_{\mit I}J_{\mit K}T $ \\ 
1834 & 4600 & --- & & $\rm12^h40^m23\fs0$ & $+03^\circ07'02''$ &  7.35 & 0.20 & $   615 $ & $ \rm T $ \\ 
1869 & 4608 & --- & & $\rm12^h41^m13\fs4$ & $+10^\circ09'18''$ & 20.23 & 0.41 & $  1633 $ & $ \rm G $ \\ 
1883 & 4612 & --- & & $\rm12^h41^m32\fs8$ & $+07^\circ18'50''$ & 14.06 & 0.41 & $  1697 $ & $ \rm G $ \\ 
1902 & 4620 & --- & & $\rm12^h41^m59\fs4$ & $+12^\circ56'32''$ & 21.28 & 0.28 & $  1012 $ & $ \rm T $ \\ 
1903 & 4621 &  59 & & $\rm12^h42^m02\fs3$ & $+11^\circ38'45''$ & 14.80 & 0.05 & $   258 $ & $ \rm GKNT $ \\ 
\hline
\end{tabular}
\end{center}
\end{table*}
\addtocounter{table}{-1}
\begin{table*}
\caption[]{Final sample of early-type galaxies (ctd)}
\begin{center}
\tabcolsep 3mm
\begin{tabular}{rcccccrcrl}
\hline
\hline
\multicolumn{3}{c}{Galaxy} & &
RA & 
Dec & 
\multicolumn{1}{c@{\hspace{2mm}}}{$D$} & 
rel. error & 
\multicolumn{1}{c@{}}{$v$} &
sources \\
\cline{1-3}
\cline{5-6}
\multicolumn{1}{c}{VCC} & 
\multicolumn{1}{c}{NGC} & 
Messier & &
\multicolumn{2}{c}{(J2000)} & 
\multicolumn{1}{c}{(Mpc)} & & 
\multicolumn{1}{c@{}}{($\rm km \,s^{-1}$)} \\
\hline
1938 & 4638 & --- & & $\rm12^h42^m47\fs5$ & $+11^\circ26'32''$ & 17.33 & 0.21 & $   956 $ & $ \rm GT $ \\ 
1939 & 4624 & --- & & $\rm12^h42^m50\fs0$ & $+02^\circ41'16''$ & 15.14 & 0.06 & $   924 $ & $ \rm GJ_{\mit I}J_{\mit K}T $ \\ 
1978 & 4649 &  60 & & $\rm12^h43^m40\fs1$ & $+11^\circ33'08''$ & 16.36 & 0.04 & $   969 $ & $ \rm GKNT $ \\ 
2000 & 4660 & --- & & $\rm12^h44^m32\fs0$ & $+11^\circ11'24''$ & 15.66 & 0.11 & $   914 $ & $ \rm GNT $ \\ 
2087 & 4733 & --- & & $\rm12^h51^m06\fs8$ & $+10^\circ54'43''$ & 14.93 & 0.18 & $   772 $ & $ \rm T $ \\ 
2092 & 4754 & --- & & $\rm12^h52^m17\fs5$ & $+11^\circ18'50''$ & 16.94 & 0.12 & $  1206 $ & $ \rm GT $ \\ 
2095 & 4762 & --- & & $\rm12^h52^m55\fs9$ & $+11^\circ13'49''$ & 10.81 & 0.41 & $   818 $ & $ \rm G $ \\ 
\hline
\end{tabular}
\end{center}
\end{table*}

\begin{figure*}[htp]
\centering
\resizebox{18cm}{!}{\includegraphics{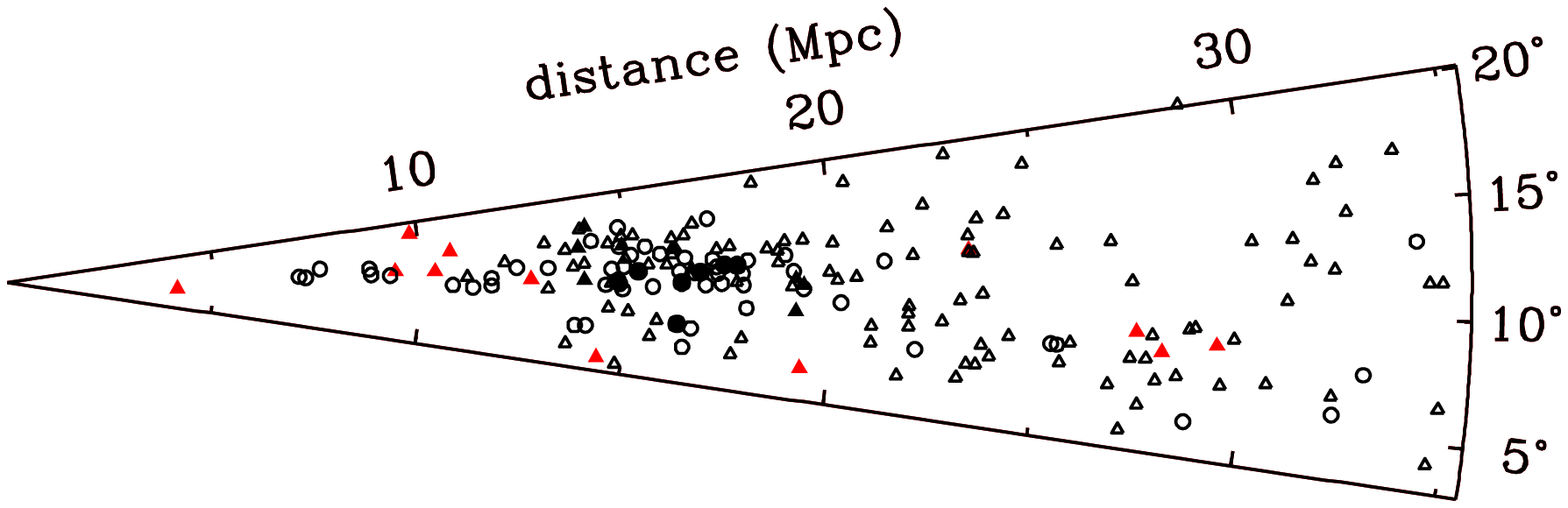}}
\resizebox{18cm}{!}{\includegraphics{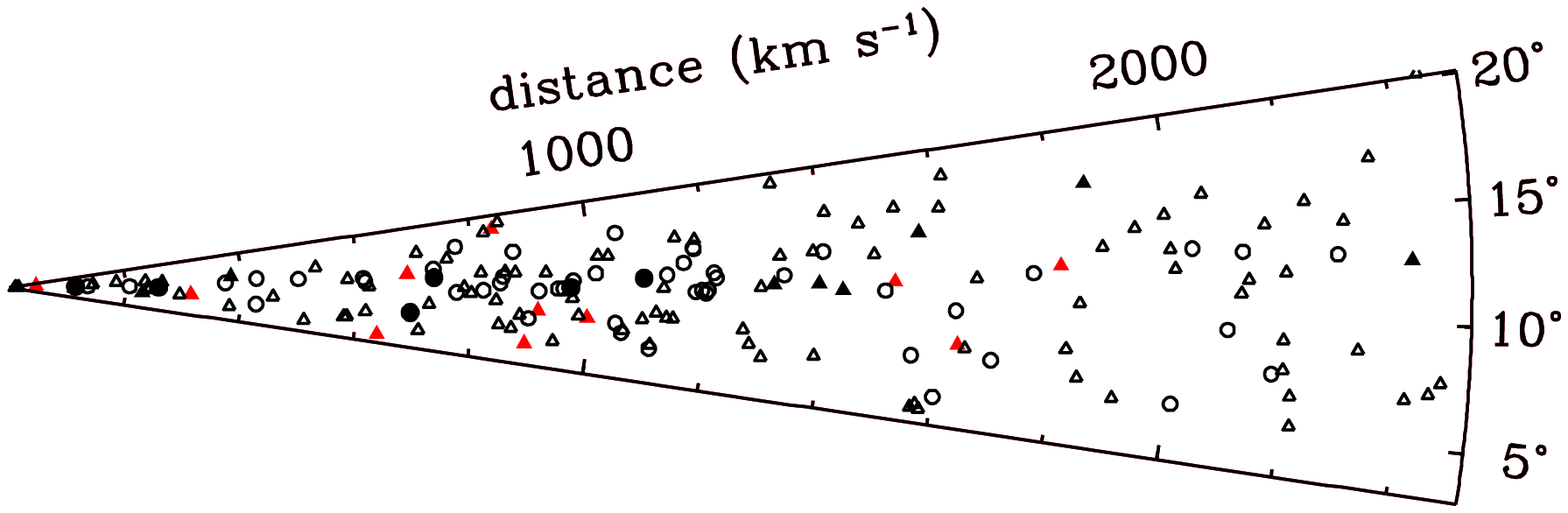}}
\caption{Positions of 146 spiral (\emph{triangles}) and 61 elliptical
(\emph{circles}) galaxies in the Virgo field (within $9^\circ$ from \object
{M87}) as
a function of Declination and
distance (assumed correct, \emph{top}) and
radial velocity (\emph{bottom}).
\HIt-deficient galaxies (deviating more than $3\sigma$ from
normalcy) are
shown as \emph{filled triangles} --- with the ones apparently outside the
cluster (Table~7, below)
shown as \emph{surrounded filled triangles} (\emph{red triangles} in
the Electronic version of the Journal), while Messier 
ellipticals (see Table~\ref{ellip}) are shown as \emph{filled circles},
with \object{M49} to the South, and \object{M59}, \object{M89}, \object{M60},
\object{M87}, \object{M86} and \object{M84} at increasing 
distances (top plot) and \object{M89}, \object{M59}, \object{M84},
\object{M60} and \object{M87} at increasing radial 
velocities (bottom plot, \object{M86} has $v < 0$).
}
\label{figwedge}
\end{figure*}

Figure~\ref{figwedge} shows a wedge diagram of the Virgo region with
the 61 early-types shown as circles and the 146 spirals shown as triangles.
One clearly notices the concentration of elliptical galaxies at a distance of
$D \simeq 17\,\rm Mpc$, which is spread out in velocity space.

\section{Observed and simulated velocity-distance relation towards the Virgo
Cluster} 
\label{replica}

For a better understanding of the kinematics of galaxies in the direction of
the Virgo cluster, we have simulated an observation of the velocity-distance
relation using the dark matter particles in the {\tt GalICS} cosmological
simulation. 

\subsection{$N$-body simulations}
\label{nbody}

The $N$-body simulations used here to find $N$-body replica of the Virgo
cluster were carried out by \cite{Ninin99}, as described by \cite{Hatton+03}
in the context of the {\tt GalICS} hybrid
$N$-body/semi-analytic model of hierarchical galaxy formation. 
Here, we are only 
interested in the density and velocity fields directly traced by dark matter
particles. The $N$-body simulation contains $256^{3}$ particles of mass $8.3
\times 10^9 M_{\odot}$ in a box of 150 Mpc size and it is run with a
softening length amounting to a spatial resolution of 29 kpc. 
The simulation was run for a flat
universe with cosmological parameters $\Omega_{0} = 0.333, \Omega_{\Lambda} =
0.667$, $H_{0}=66.7\,\rm km\,s^{-1}\,Mpc^{-1}$, and $\sigma_8 = 0.88$. 
Once the simulation is run,
halos of dark matter are detected with a `Friends-of-Friends' (FoF) algorithm
\citep{DEFW85},
with a variable linking length such that
the minimum mass of the FoF groups is $1.65 \times 10^{11} M_{\odot}$ (20
particles) at any time step. With this method, over $2 \times 10^{4}$ halos are
detected at the final timestep, corresponding to the present-day ($z=0$)
Universe.
The {\tt GalICS} halo finder does not allow halos within halos, so that a
cluster, to which is assigned a massive halo, cannot contain smaller halos
within it.

\subsection{Picking isolated halos}
\label{isolated}

Among the 12 halos that {\tt GalICS} produces at $z=0$ with $M_\mathrm{FoF} >
10^{14}\, M_\odot$, we wish to choose those that resemble the Virgo cluster,
both in terms of mass and environment.

As we can see in the radial phase space diagrams of dark matter halos
of the simulations (Fig.~1 of \citealp{MSSS03}),
it is quite common to find small groups around the main
cluster halo, whereas the Virgo cluster is  not known to have massive
neighbors, 
except for the \object{M49} 
group just within the virial radius.   
Note that the size of the simulation box, $L=150\,\rm Mpc$,
implies, 
given the periodic boundary condition, a maximum separation between halos of
$\sqrt{3}\,(L/2)=130\,\rm Mpc$.

The effect of a neighboring halo will be from its tide on the
test halo. The ratio of the tidal acceleration on a particle of the test halo
at distance $r$ from its center caused by a neighboring halo at distance $R$
from the center of the first halo
to the acceleration of this particle by the potential of its halo amounts to
a mean density criterion:
\begin{equation}
a_\mathrm{tid} \approx {G\,M_2(R) \over R^3}\,r
\end{equation}
and
\begin{equation}
a_\mathrm{halo} = {G\,M_1(r) \over r^2}
\end{equation}
so that 
\begin{equation}
{a_\mathrm{tid} \over a_\mathrm{halo}} \approx {\rho_2(R) \over \rho_1(r)} \ ,
\end{equation}
where the densities are average (and not local) and subscripts 1 and 2 stand for
the central and the neighboring halo respectively.
For clarity, we denote $r_1$, $M_1$, $r_2$ and $M_2$ the virial radii and masses for the
test and neighboring halo, respectively (i.e. we drop the `100' subscript).
Then, 
equating $r = r_1$, noting that the mean density at the virial
radius is the same for each halo, i.e. 100 times the critical density of the
Universe, and writing 
\begin{equation}
\eta = {R\over r_2} \ ,
\label{eta}
\end{equation}
the influence of each neighbor is measured by the mean density of the
neighbor measured at the center of the test halo
\begin{eqnarray}
\rho_2(R) &=& {3\,M_2(R)\over 4\,\pi\,R^3} \nonumber \\
&\propto& {M_2(R) / M_2 \over \eta^3}
= \left \{\eta^3 \left [ \ln \eta - \eta/(1+\eta) \right ] \right\}^{-1} \,,
\label{tidecrit}
\end{eqnarray}
where we assumed NFW density profiles and
where the last equality is from \cite{CL96}.
Since the mean density is a monotonously decreasing function of radius (i.e. 
the function of $\eta$ in eq.~[\ref{tidecrit}] decreases with $\eta$), the
most isolated halos will have the highest minimum values of $\eta$.

The output from {\tt GalICS} includes the mass of each halo obtained from the
FoF estimator, so that one can easily perform a first guess of which of the
halos with mass close to that of Virgo are as isolated as appears the Virgo
cluster. 
Table~3
lists for the 10 most massive halos in the
simulation, the parameters of its most perturbing neighbor (that with the
lowest value of $\eta$) using FoF quantities for radii and masses. 
Columns 2 to 4 list the virial radius, mass and circular velocity obtained
using the more precise spherical overdensity (SO) method, which allow a
direct comparison with the value we obtained for the Virgo cluster 
in \citet{MSSS03}.

\begin{table}[ht] 
\begin{center}
\caption{Most perturbing neighbors of 10 most 
massive halos in {\tt GalICS} simulation}
\tabcolsep 1.5mm
\begin{tabular}{r@{\ \ \ \ }cccccc}
\hline
\hline
\multicolumn{1}{c}{rank} 
& $r_1^\mathrm{SO}$ & $M_1^\mathrm{SO}$ & $v_1^\mathrm{SO}$ & $R$ 
& $M_2^\mathrm{FoF}$ 
& $\eta_\mathrm{min}$ \\
\multicolumn{1}{c}{(FoF)} 
& (Mpc) & ($10^{14}\,M_\odot$) & ($\rm km \, s^{-1}$) 
& $\overline{r_1^\mathrm{SO}}$
& $\overline{M_1^\mathrm{FoF}}$ 
 \\
\hline
1    & 3.19      & 16.7~\,   & 1503       & 0.55 & 0.011	&3.7\\
2    & 2.42      &  7.4      & 1143       & 0.58 & 0.003	&3.1\\     
3    & 2.50      &  8.1      & 1180       & 0.55 & 0.020	&2.4\\      
4    & 2.54      &  8.5      & 1200       & 0.53 & 0.011	&2.4\\ 
5    & 2.10      &  4.8      &~\,989      & 0.62 & 0.001	&2.9\\ 
6    & 2.00      &  4.1      &~\,942      & 0.58 & 0.001	&2.5 \\ 
7    & 1.97      &  3.9      &~\,927      & 0.57 & 0.031	&3.4\\ 
8    & 1.78      &  2.9      &~\,842      & 0.61 & 0.002	&2.5\\ 
9    & 1.84      &  3.2      &~\,869      & 0.58 & 0.103	&2.2\\ 
10   & 2.03      &  4.3      &~\,958      & 0.51 & 0.003	&2.6 \\ 

\hline
\end{tabular}
\end{center}
\label{isolatedFoF}
\end{table}
As seen in 
Table~3, 
the most perturbing halo is always inside $r_1$.
Therefore, it is not clear
that we can easily estimate
the exact virial radius and mass of each perturbing halo
with the same spherical overdensity method as in \citet{MSSS03} to
produce a better estimate of isolated halos. As we expect $r_1^\mathrm{SO}$
to be 
approximately proportional to $r_1^\mathrm{FoF}$,
we take the results of 
Table~3
as good indicators of the degree of isolation of halos in the
simulations.

As shown in 
Table~3, 
the most isolated halo in the
simulations is 
halo 1, the most massive halo. 
Picking the most massive halo also ensures that no other massive halos will
distort the mock velocity-distance relation.

\subsection{Rescaling the simulations to the scales of the Virgo cluster}

We now need to rescale 
the simulated clusters to Virgo.
A reasonable way to put two halos in
comparable units is to normalize distances with the virial radius,
$r_{100}$, and velocities
with the circular velocity of the halo at the virial radius, $v_{100}$. 
We therefore convert from the simulation frame to the
Virgo cluster frame by rescaling the terms related to distance with the
factor (omitting the `100' subscripts for clarity)
$r_\mathrm{V}/r_\mathrm{s}$
and the terms related to
velocities by the analogous factor
$v_\mathrm{V}/v_\mathrm{s}$,
where superscripts `s' and `V' refer to the
simulation and the Virgo cluster, respectively.
We adopt the Virgo cluster 
virial radius and circular velocity derived by \cite{MSSS03} from X-ray
observations
$r_\mathrm{V} = 1.65\,h_{2/3}^{-1}\,\rm Mpc$, where 
$h_{2/3} = H_0 / (66.7 \, \rm km \, s^{-1} \,Mpc^{-1})$, and
$v_\mathrm{V} = 780 \, \rm km \, s^{-1}$.

Moreover, the $N$-body simulations give
us comoving velocities ${\bf u}_\mathrm{s}$ of dark matter particles, to
which we must add
the Hubble flow.
Therefore:
\begin{eqnarray}
\label{d}  
D & = & D_\mathrm{s} \,\frac{r_\mathrm{V}}{r_\mathrm{s}}  \\
\label{v}
v & = & u_\mathrm{s} \, \frac{v_\mathrm{V}}{v_\mathrm{s}} + H_\mathrm{0}\, D  
\end{eqnarray}
where $u_\mathrm{s}$ represents the comoving radial velocity of the particle
in the simulation with respect 
to the observer in comoving units.

Inverting equation~(\ref{d}), we need to place the observer at a distance
\begin{equation}
D_\mathrm{obs} = D_\mathrm{V} ~\frac{r_\mathrm{s}}{r_\mathrm{V}}
\label{dobs}
\end{equation}
to the halo used to mimic the Virgo cluster (in simulation units).
Once we have $D_\mathrm{obs}$, we can select different observers by choosing
different positions on a sphere of radius $D_\mathrm{obs}$ centered on the
halo and assigning to each position, 
the mean velocity of the 10 nearest
halos to the observer, 
so as to give the observer an appropriate peculiar velocity.
As our galaxies in the Virgo cluster extend up to $D_\mathrm{max} = 50\,\rm
Mpc$,  we will need
to use particles around the chosen dark matter halo up to a distance from the
cluster of $D_\mathrm{max} - D_\mathrm{V} \simeq 20\, r_{100}$.

It is interesting to check the peculiar velocities (or equivalently the
systemic velocities) found for the simulated
observer relative to the simulated cluster and compare to that of our Local
Group of galaxies relative to the Virgo cluster, after rescaling the
simulation radial and velocity separations to the Virgo cluster attributes as
in equations~(\ref{d}) and (\ref{v}).
For 500 observers placed at a given distance from the halo, but in different
random directions, we find a systemic velocity between the halo and the
observer (i.e. the mean of
10 nearest halos) 
in Virgo units of $998\pm118 \, \rm km \, s^{-1}$ (errors are the $1\,\sigma$
dispersion) when we assume
$D_\mathrm{V} = 16.8 \,\rm Mpc$ (as suggested by the positions of the
early-type 
galaxies in Fig.~\ref{figwedge}, and especially of \object{M87},
which lies at the center of the X-ray emission) 
and $1308\pm155 \, \rm km \, s^{-1}$ when we
assume $D_\mathrm{V} = 21 \,\rm Mpc$, as in \cite{Sanchis+02}.
The values of the systemic velocity of the Virgo cluster in
the literature range approximately between 900 and 1000~$\rm km~\rm s^{-1}$ 
\citep{TBGP92}. Therefore, putting the Virgo cluster at 16.8 Mpc, as indicated by the early type galaxies,
gives a better match to the systemic velocity of the cluster.
In addition, if we calculate the number of all galaxies (early-type and
late-type ones) per distance bin, we found the maximum of the distribution
between $16-18\,\rm Mpc$. Even if our sample lacks of completeness, this
result, 
together with the previous ones, lead us to choose $D_\mathrm{V}=16.8 \rm
\,Mpc$ 
for the comparison between the $N$-body simulations and the Virgo cluster.

\subsection{Velocity-distance relation without distance errors} 

We take all the objects lying within a cone of half-opening $9^\circ$ 
aligned with the axis connecting the observer to the center of the halo.

Because observed galaxy catalogs, such as the VCC \citep*{BST85} catalog of
Virgo cluster galaxies, are magnitude-limited, we imposed upon our simulated
catalog (of dark matter particles) the same magnitude limit as the VCC,
namely $B < 18$ \citep{BST85}.
For this, we consider bins of distance to the observer, and, in each distance
bin, we remove a fraction of dark matter particles equal to the predicted
fraction of particles fainter than $B = 18$ 
given the distance and the gaussian-shaped
luminosity function of spiral galaxies derived by
\cite*{SBT85}.
This method supposes that the luminosity function is independent of
environment (i.e. distance to the center of the Virgo cluster) and that those
VCC 
galaxies with redshift-independent distance measurements have the same
magnitude limit as the VCC galaxies in general.
This luminosity incompleteness is only noticeable at large
distances, where it amounts to reducing
the number of particles by less
than 10\%.

\begin{table*}
\label{refused}
\begin{center}
\caption{Spiral galaxies without \HIt\ data}
\begin{tabular}{cccccccc}
\hline \hline
\multicolumn{2}{c}{Galaxy} & & \multicolumn{1}{c@{}}{RA} 
&\multicolumn{1}{c@{~~}}{Dec}  & $T$ &\multicolumn{1}{c@{~}}{$v$}
&\multicolumn{1}{c} {$D$}       \\ 
\cline{1-2}
\cline{4-5}
\multicolumn{1}{c}{VCC} & \multicolumn{1}{c}{NGC}   & &
\multicolumn{2}{c@{}}{(J2000)}  &     &\multicolumn{1}{c@{~}}{(km~s$^{-1}$)} &\multicolumn{1}{c} {(Mpc)}    \\ 
\hline
~\,132& --- & & $ \rm 12^h15^m03\fs9$ &  $+13^\circ01'55''$ &8&   ~1965 &
$10.76^{+2.60}_{-2.10}$  \\ 
~\,343& --- & & $ \rm 12^h19^m22\fs0$ &  $+07^\circ52'16''$  &8&   ~2336 &
23.77~~~~~~        \\ 
~\,912& 4413 & & $ \rm 12^h26^m32\fs2$ &  $+12^\circ36'39''$ &2&    ~\,$-$13 &  $15.63^{+2.57}_{-2.20}$  \\
~\,952& --- & & $ \rm 12^h26^m55\fs5$ &  $+09^\circ52'58''$ &5& ~~\,853 &   $~\,4.77^{+4.60}_{-3.88}$  \\
1605& --- & & $ \rm 12^h35^m14\fs5$ &  $+10^\circ25'53''$ &5& ~~\,951 &
$24.66^{+6.82}_{-5.34}$  \\ 
1673& 4567 & & $ \rm 12^h36^m32\fs9$ &  $+11^\circ15'28''$ &4&   ~2133 &
26.30~~~~~~         \\ 
1933& --- & & $ \rm 12^h42^m44\fs7$ &  $+07^\circ20'16''$ &2&   ~2271 &
$35.16^{+0.33}_{-0.32}$  \\ 

\hline
\end{tabular}
\end{center}
\end{table*}       

\begin{figure*}
   \centering
   \includegraphics{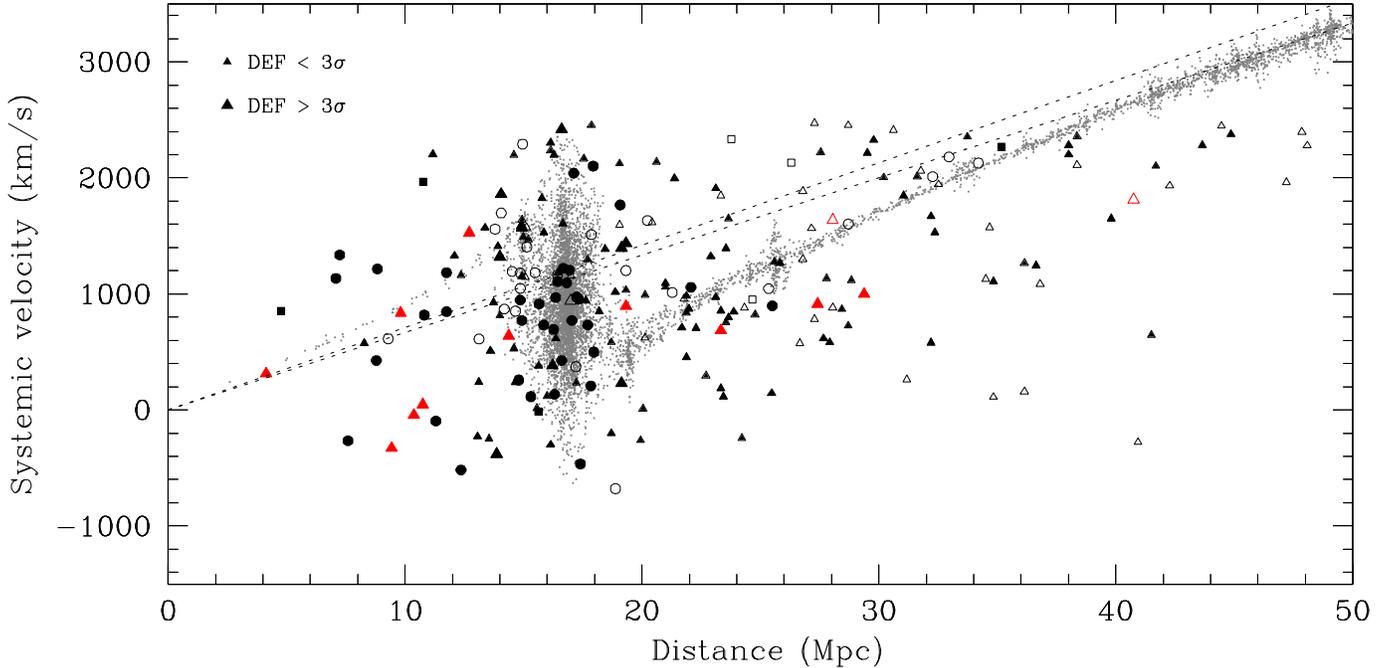}
      \caption{Simulated and observed velocity-distance diagrams. The
   observer is at $(0,0)$ and sees a cone of angular radius $9^\circ$.
\emph{Dots} represent the velocity field traced by the particles (1 in 50 for
   clarity) in the
   cosmological $N$-body simulation in the direction of the most massive
   halo, for which 
distances and velocities have been rescaled (eqs.~[\ref{d}] and
   [\ref{v}]) incorporating a Hubble flow with $H_0 = 66.7 \,\rm km \,s^{-1} \,
   Mpc^{-1}$, as in the simulation. Superposed are galaxies of the Virgo
   cluster. 
  \emph{Circles} 
   and \emph{triangles}
   represent 
   early-type and late-type galaxies, respectively. The size of the triangles
   is proportional to the H$\,${\scriptsize I}-deficiency of the spiral  
	     galaxies measured in units of the mean standard deviation for
   field objects (= 0.24). 
   \emph{Squares} represent additional
	     spiral galaxies with no \HIt-deficiency data.
	     \emph{Open symbols} indicate galaxies with uncertain distances
   (with distance errors greater than 5 Mpc or spirals with only
   one distance estimate). 
\emph{Dashed} lines show the unperturbed Hubble flow with $H_0 = 66.7$ and
   $70 \,\rm km \,s^{-1} \, Mpc^{-1}$, respectively (going upwards).
The locations of the vertical spikes of simulated particles
away from the Virgo cluster  (assumed to lie at 16.8 Mpc) are not meaningful
   and should not be compared to the measured galaxy distances.
}
         \label{proj}
\end{figure*}

Figure~\ref{proj} shows the velocity-distance plot of the simulated dark
matter particles (\emph{small points}), rescaled to Virgo units, and from
which particles 
statistically fainter than the magnitude limit of the observed catalogs were removed.
Superposed are the spiral galaxies (\emph{triangles}) from \cite{Solanes+02}
and the early-type galaxies (\emph{circles}) compiled in
Table~\ref{ellip}. 
We 
have added to our sample the spiral galaxies compiled by
\cite{Solanes+02} with reliable Tully-Fisher
(\citealp{TF77}, hereafter TF) distances but no \HI\ data.
We list these galaxies in 
Table~4, 
with their
coordinates, morphological type (from {\sf LEDA}), systemic
velocity 
and the final 
distance with errors assigned to it by \cite{Solanes+02}, who estimated
errors in distance as the mean deviation between the rescaled mean distance
and the different estimates used (with this method,
galaxies with only one estimator cannot be assigned an error and
their distances are considered uncertain).  
We represent these objects as
\emph{squares} in Figure~\ref{proj}.

Although Figure~\ref{proj} concerns a single observer and a single halo to
represent the Virgo cluster (i.e. halo 1, which is the most isolated), 
the results are very similar for all halos studied as well as for any
observer whose velocity with respect to the center of the main halo is similar
to the mean of the 500 random trials.
In other words, Figure~\ref{proj} provides a realistic general
velocity-distance relation for a Virgo-like cluster in a flat Universe with a
cosmological constant.
Very recently, \cite{KHKG03} have produced a similar plot for dark matter
particles from their adaptive mesh refinement simulation.
Note that the groups visible in the 3D phase space plot (Fig.~1 in
\citealp{MSSS03})
are much less visible 
here and the projection places them at smaller distances from the center of
the main cluster.

Figure~\ref{proj} indicates that the velocity field drawn by the observed
spirals is noticeably different from the one followed by the dark matter
particles of the simulation. Nevertheless, the figure
shows that 1)  the center of
the halo corresponds to the location of greater galaxy density 
(not only for the early-type galaxies used to mark the center of the Virgo
cluster), 2) the 
thickness of the halo in velocity space (the velocity dispersion) 
corresponds to the corresponding thickness of the
central observed galaxies, and 3) the majority of the elliptical galaxies outside the cluster
also 
appear to fit well the location of the particles in the outer
infalling/expanding locus.
These 3 facts suggest that both the
distance to the halo and the rescaling factors are well chosen and that the
differences between the Virgo cluster and the halo in the simulation should be
attributable to errors arising from the distance estimates (as the velocity errors
are  negligible). It appears unlikely that the galaxies that, for a given
distance, lie within more than $1000 \, \rm km \, s^{-1}$ from the
locus of dark matter particles belong to the low- or high-velocity tails of
large velocity dispersion groups near the cluster, since in our simulations,
there are no such high velocity dispersion groups outside the main cluster.

\subsection{Velocity-distance relation with distance errors} 
\label{role}

\begin{figure*}
   \centering
  \includegraphics{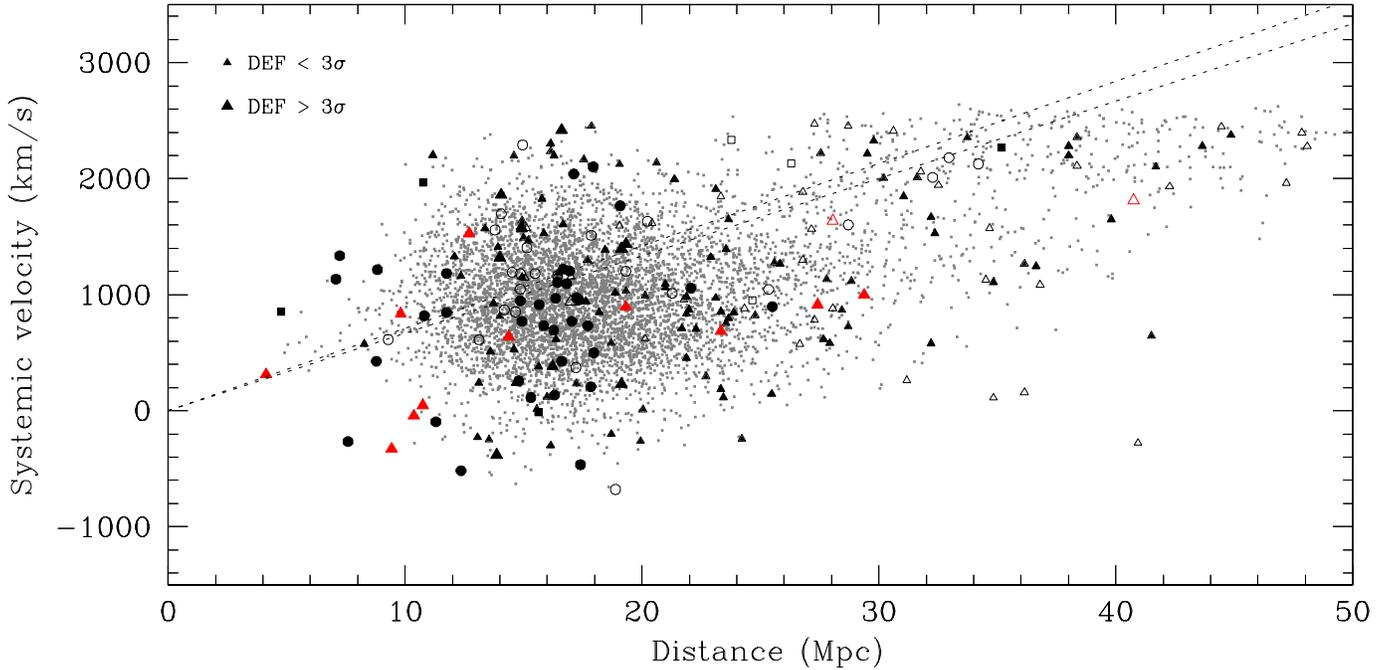}
      \caption{Same as Figure 2, but incorporating gaussian relative
   distance errors of $\sigma \ln D = 0.2$ for the dark matter particles.}
          \label{errors}
   \end{figure*}

Given that most of the galaxies outside the velocity field drawn by dark matter
particles are spirals, we now build
a new velocity-distance diagram of our simulated particles incorporating the
mean relative distance error for the spiral galaxy sample.
\cite{Solanes+02} combined different studies of the TF relation to calculate
the distances of 
spirals in the Virgo cluster. These studies give dispersions of the TF
relation of roughly 0.4 mag., corresponding to an uncertainty
of 18\% in relative distance. 
Figure~\ref{errors} shows the corresponding velocity-distance plot for our
simulated Virgo line of sight, with inclusion of gaussian relative distance
errors with $\sigma \ln D = 0.20$ (a slightly more conservative value than
0.18). 
We omit particles beyond 40 Mpc, which corresponds to the distance (before
folding in the distance errors) where the 
particles have a systemic velocity approximately equal to the maximum of the observed
galaxy sample.

Figure~\ref{errors} indicates that the inclusion of distance errors, at the
20\% relative rms level, allows us
to reproduce fairly well the observed velocity-distance diagram.
There are still several galaxies that lie far outside the general locus of
the dark matter particles. Among them are two galaxies 
(\object{VCC 319}, which is part of a galaxy
pair and \object{M90} = \object{NGC 4569} = \object{Arp 76}, which is tidally
perturbed by a close neighbor) apparently in the foreground of the cluster
between 7 and 10 Mpc and with $v < -200 \, \rm km \, s^{-1}$, and four
galaxies (IC 3094,
\object{NGC 4299}, which is part of a pair, \object{NGC 4253} and \object{VCC
1644}) in the background between 34 and 42 Mpc and $v < 800 \, \rm km \,
s^{-1}$.  Inspection of Figure~\ref{proj} indicates that these four 
galaxies are
at more than $2000 \, \rm km \, s^{-1}$ from the expected locus of the
Tolman-Bondi pattern given negligible distance errors. It is highly unlikely
that the peculiar velocities in interacting pairs can reach values as high as 
$2000 \, \rm km \, s^{-1}$,
hence the distances to these galaxies appear to be grossly incorrect.

We list the most discrepant galaxies in 
Table~5.
\begin{table}[ht]
\begin{center}
\caption{Grossly incorrect galaxy distances}
\tabcolsep 1mm
\begin{tabular}{rcccccr@{\ \ \ \ }r}
\hline
\hline
\multicolumn{2}{c}{Galaxy} & & RA & Dec & \multicolumn{1}{c}{$T$} 
& \multicolumn{1}{c}{$v$} & 
\multicolumn{1}{c}{$D$} \\
\cline{1-2}
\cline{4-5}
VCC & NGC & & \multicolumn{2}{c}{(J2000)} & & 
\multicolumn{1}{c}{($\rm km \, s^{-1}$)} & Mpc \\
\hline
213& ---& & $\rm 12^h16^m56\fs0$ & $+13^\circ37'33''$& 5 & $-278$&40.9\\ 
319 & ---& & $\rm 12^h19^m01\fs7$ & $+13^\circ58'56''$& $-4~~\,$ &$-256$ & 7.6 \\
491 &4299 & & $\rm 12^h21^m40\fs9$ & $+11^\circ30'03''$ & 8 &112&34.8  \\ 
1524& 4523 & & $\rm 12^h33^m48\fs0$ & $+15^\circ10'05''$ & 9 &158&36.1\\ 
1644& & & $\rm 12^h35^m51\fs8$& $+13^\circ51'33''$ & 9 &646&41.5   \\ 
1690& 4569 & & $\rm 12^h36^m49\fs8$& $+13^\circ09'48''$ & 2 &$-328$& 9.4   \\ 
\hline
\end{tabular}
\end{center}
VCC 1690 = NGC 4569 = M90 = Arp 76.
\label{baddists}
\end{table}
Most of these galaxies have a single distance measurement (the elliptical
\object{VCC 319} has a single S\'ersic fit distance estimator, which is known
to be inaccurate --- see Table 1).
However, it is a surprise to find a bright object such as \object{NGC~4569}
(\object{M90}) in this list, especially since it has 6 different
Tully-Fisher distance measurements, 5 of which yield a distance between 7.9
and 10.2 Mpc and one \citep{Gavazzi+99} giving $D = 16.1\,\rm Mpc$, which is
what we expect from Figure~\ref{proj}.

Since the 20\% relative errors correctly reproduce the envelope of particles
from the cosmological simulation, increasing the relative rms error to say
30\% 
will not explain
the distances to these discrepant spirals.
We therefore conclude that the distances of the galaxies listed in 
Table~5
must be highly inaccurate.

In their dynamical, Tolman-Bondi, modeling of the Virgo region, 
\cite{Sanchis+02} only used those spiral galaxies believed to trace the outer
envelope of galaxies expanding out of the Virgo cluster.
Figure~\ref{proj} clearly indicates that the galaxies used by
\citeauthor{Sanchis+02} (roughly aligned between distances of 34 and 45 Mpc) lie at typically $600 \, \rm km \, s^{-1}$ lower
velocities for their distance than predicted by our simulated velocity-distance
model. Figure~\ref{errors} suggests that the alignment of these galaxies
in the velocity-distance plot
may be fortuitous, given the $\sim 20\%$ relative distance errors.
This fortuitous alignment forced the model of \citeauthor{Sanchis+02}
to
shift the center of the Virgo cluster to greater distances. 

\subsection{Tolman-Bondi estimates of the Virgo cluster mass and distance}
We now use the cosmological $N$-body simulations to quantify the errors on
the distance and mass to 
the Virgo cluster obtained by fitting a Tolman-Bondi model to the
velocity-distance data outside the cluster core.
We construct simulated pseudo-galaxy samples 
by randomly selecting as many points from the cosmological $N$-body 
simulation as there are galaxies in the Virgo
cluster sample. We apply fixed relative distance errors in the range 
0\%--30\% and, for each relative distance
error, we construct 50 realizations of the Virgo sample. We identify
pseudo-galaxies 
presumably belonging to the core by building histograms of the number
of pseudo-galaxies 
per 2 Mpc distance bin. Such histograms peak at the core of the
halo,
in a region of width  $\sim 4$~Mpc for the no error set and up
to $\sim12$~Mpc for
the 30\% 
error set.
We avoid the virialized core by only considering pseudo-galaxies 
lying on the back side of the identified cluster core. 

In the Tolman-Bondi 
model used by \citeauthor{Sanchis+02}, galaxies are
treated as test particles  
moving in a point-mass gravitational potential that is independent of time.
The model velocity-distance curve is a function of the age of the Universe,
$t_0$, the distance to the cluster, $D_\mathrm{V}$, the point mass of the
cluster, 
$M_\mathrm{V}$, and the systemic velocity of the cluster relative to the
Local Group 
of galaxies, $v_\mathrm{V}$.
Only three of these four parameters are independent, but
given the poor current velocity-distance data, it is better to use only two
free parameters and fix the third one. 
Following \citeauthor{Sanchis+02}, we
have fixed $v_\mathrm{V}$ to the mean velocity of the main 
halo for the 50 realizations (998~$\rm km~\rm s^{-1}$ for a distance of
16.8~Mpc, 
see section 3) and we have allowed $M_\mathrm{V}$ and $D_\mathrm{V}$ to vary
freely. The resulting
$t_0$ will help us to see the viability of the results. 

Table~6 shows the results of the fits to the point-mass
Tolman-Bondi relation (see \citealp{Sanchis+02}) for the different sets.  
For zero distance errors, the model
reproduces fairly well the input data of  $D_\mathrm{V} = 16.8\,\rm Mpc$ and
$M_\mathrm{V} = 10^{14.35}\, M_\odot$ \citep{MSSS03}, as  well as the age of
the Universe: $t_0 = 13.7 \,\rm Gyr$ from the {\tt GalICS} cosmological
simulations (which happens to perfectly match the age recently obtained by
\citealp{Spergel+03} from the {\sf WMAP} CMB experiment). The Table also
clearly indicates that the errors in distance contribute to a slightly larger
distance to the cluster (by up to 2\%), combined with a much larger mass for
the cluster (up to a factor 40!). 
This dramatic rise in the mass decreases the resulting $t_0$ to unacceptably
low values. Therefore, distance errors can cause the fitting of
Tolman-Bondi solutions to the velocity vs. distance of galaxies to yield
incorrect cluster parameters.


\begin{table}
\label{model_san}
\caption{Best fitting parameters for each dataset}
\begin{center}
\begin{tabular}{ccccccc}
\hline 
\hline
Error  & \multicolumn{2}{c}{$D_\mathrm{V}$} & &
\multicolumn{2}{c}{$\log(M_\mathrm{V})$}  &  $t_0$  \\
       & \multicolumn{2}{c}{(Mpc)}   & & 
\multicolumn{2}{c}{(M$_{\odot}$)}            &   (Gyr)        \\  
\cline{2-3}
\cline{5-6}
       &   mean & $\sigma$  & &  mean &  $\sigma$  &  \\ \hline
~\,0\% & 16.7   & 0.5  &  &  14.5 &   0.2 &  14.1\\
~\,5\% & 16.8   & 0.5  &  &  14.6 &   0.3 &  13.9\\
10\%   & 16.9   & 0.6  &  &  14.9 &   0.4 &  12.8\\
15\%   & 17.0   & 0.8  &  &  15.1 &   0.5 &  12.1\\
20\%   & 17.1   & 0.9  &  &  15.5 &   0.5 &  10.1 \\
25\%   & 17.1   & 1.0  &  &  15.7 &   0.4 &  ~\,9.0 \\
30\%   & 17.1   & 1.0  &  &  16.1 &   0.5 &  ~\,6.7  \\
\hline

\end{tabular}
\end{center}
\end{table}
For the case
of real data on the Virgo cluster, the same fitting procedure leads to
$M_\mathrm{V} 
\sim 3\times 10^{16}\, M_{\odot} $ and $D_\mathrm{V} \sim 17.4 \rm\, Mpc$, 
which imply $t_{0} = 5.2$ Gyr. 
Note that the present fitting procedure takes into account all galaxies
beyond the identified core of the cluster, whereas \citeauthor{Sanchis+02}
only used those galaxies that are aligned between distances of 34 and 45 Mpc
in the velocity-distance plot, deducing a very different distance to the Virgo cluster (between 20 and 21
Mpc). 
The comparison of the values of $t_0$ from the real data and from the
simulated data convolved with different relative distance errors, suggests
that the TF distances may have up to 30\% relative errors.

As shown above, the estimation of the {\em distance} to the cluster provided
by the point mass model is relatively good even for appreciable errors in
relative distances. Therefore, one could still obtain a reasonable value for
the mass of the cluster by taking $D_\mathrm{V}$ from the fit (i.e., 17.4 Mpc),
together with $v_\mathrm{V} = 980~\rm km~\rm s^{-1} $ 
\citep{TBGP92} and $t_0 = 13.7$~Gyr \citep{Spergel+03}
as input parameters for the model and computing $M_\mathrm{V}$ as a
result. In this 
way, we obtain a value of $M_\mathrm{V}= 7 \times 10^{14}\,M_{\odot}$. Notice
that the 
effective mass of the point mass model does not have to coincide with the
virial mass of the cluster, as the latter one represents only the virialized
part of the cluster, while the former gives us an idea of the dynamical mass
of the whole Virgo region.

\section{Discussion} \label{deficiency}

As mentioned in Sec.~1 and clearly seen in Figure~\ref{proj}, there is a
non negligible number of \HI-deficient galaxies distant from the cluster
core, in particular at distances of 10 and 28 Mpc from the Local Group.
We now examine different explanations for the presence of these
\HI-deficient
galaxies outside the core of Virgo:

\begin{enumerate}
\item very inaccurate distances (perhaps from the inadequacy of the Tully
Fisher relation for galaxies undergoing ram pressure or tidal stripping) for
galaxies that are in fact within the 
Virgo cluster (in the core or more probably in a rebound orbit) and which
would have lost some of their gas by ram pressure 
stripping from the intracluster gas;   
\item tidal interactions from members of a same group;
\item tidal interactions from neighbors;
\item recent minor mergers heating the gas;
\item overestimated \HI-deficiency in early-type spirals.
\end{enumerate}

In 
Table~7 
are listed all galaxies from the compilation of
\cite{Solanes+02} which satisfy at least one of the following criteria:
\vbox{
\begin{itemize}
\itemsep 3pt
\item[$\bullet$] $D < 13 \,{\rm Mpc}$ ,
\item[$\bullet$] $ 21 < D < 50 \,{\rm Mpc}$ ,
\item[$\bullet$] $\theta / \theta_{100} > 1$ ,
\end{itemize}
}
where $\theta$ is the angular distance to \object{M87} and $\theta_{100} =
r_{100}/D_\mathrm{V} = 5\fdg6$ is the angular virial radius of the Virgo
cluster.
Columns (1) and (2) give the galaxy names, 
the coordinates in columns (3) and (4), 
the galaxy type, total blue extinction corrected magnitude and optical
diameter (all three from {\sf LEDA}) are respectively in
columns (5), (6), and (7),
the projected distance to M87 in
units of the virial radius ($r_{100}$) of the cluster in column (8), 
the velocity relative to the Local Group taken from the Arecibo General Catalog
(a private dataset maintained by R. Giovanelli and M. P. Haynes at the
University of Cornell), 
in column (9) 
and column (10) gives the adopted distance \citep{Solanes+02}.
The last 3 columns of
Table~7
are explained in Sect.~\ref{wrongdist}.
\begin{table*}[ht]
\label{outliers}
\caption{\HIt-deficient galaxies possibly outside the Virgo cluster}
\begin{center}
\begin{tabular}{cccccccccr@{\ \ \ \ \ }crcc}
\hline
\hline
\multicolumn{2}{c}{Galaxy} & & RA & Dec & $T$ & $B_T^c$ & 
$D_{25}$ & $\theta$ & \multicolumn{1}{c}{$v$} & $D$ & \multicolumn{1}{c}{\
\ $N$} & $P_1$ & $P_{2.5}$ \\ 
\cline{1-2}
\cline{4-5}
\multicolumn{1}{c}{VCC} & NGC & & \multicolumn{2}{c}{(J2000)} & & & &
& \multicolumn{1}{c}{ ($\rm km \, s^{-1}$)} & (Mpc) \\
\hline
--- & 4064  & & $\rm 12^h04^m11\fs8$ & $\rm 18^\circ26'33''$ & 1 & 11.73 &
    $4\farcm0$& $8\fdg8$ & 837 & \, 9.8 & 942 & 0.00 & 0.81 \\
522 & 4305 & & $\rm 12^h22^m03\fs5$ & $\rm 12^\circ44'27''$ & 1 & 12.87 &
$2\farcm0$ &  $2\fdg2$ & 1814 & 40.7 & 200 & 0.00 & 0.00 \\ 
524 & 4307 & & $\rm 12^h22^m06\fs3$ & $\rm  09^\circ02'27''$ & 3 & 11.84 &
$3\farcm5$ &  $4\fdg0$ & 913 & 27.4 & 1846 & 0.41 & 0.48 \\
559 &  4312 & & $\rm 12^h22^m32\fs0$ & $\rm 15^\circ32'20''$ & 2 & 11.76 &
$4\farcm7$ & $3\fdg7$ & 47 & 10.8 & 754 & 0.95 & 1.00 \\
713 & 4356 & & $\rm 12^h24^m15\fs9$ & $\rm  08^\circ32'10''$ & 6 & 13.02 &
$2\farcm6$ & $4\fdg2$ & 998 & 29.4 & 1377 & 0.19 & 0.23 \\
979 & 4424 & & $\rm 12h27^m13\fs3$ & $\rm  09^\circ25'13''$ & 1 & 11.99 &
$3\farcm4$ & $3\fdg1$ &314 & \, 4.1 & 106 & 0.00 & 0.00 \\
1043 & 4438 & & $\rm 12^h27^m46\fs3$ & $\rm 13^\circ00'30''$ & 1 & 10.55 &
$8\farcm7$ & $1\fdg0$ & --45 & 10.4 & 755 & 1.00 & 1.00 \\
1330 & 4492 & & $\rm 12^h30^m58\fs9$ & $\rm  08^\circ04'41''$ & 1 & 13.04 &
$1\farcm9$ & $4\fdg3$ & 1638 & 28.1 & 1261 & 0.08 & 0.11 \\
1569 &--- & & $\rm 12^h34^m31\fs4$ & $\rm 13^\circ30'23''$ & 5 & 14.62 &
$0\farcm8$ & $1\fdg4$ & 687 & 23.3 &  8967 & 0.72 & 0.94\\ 
1690 & \ \,4569$^a$ & & $\rm 12^h36^m50\fs5$ & $\rm 13^\circ09'54''$ & 2 & \ \,9.63 &
$10\farcm4$\ \  & $1\fdg7$ &--328 &  \, 9.4 & 143 & 1.00 & 1.00 \\
1730 & 4580 & & $\rm 12^h37^m49\fs5$ & $\rm  05^\circ22'09''$ & 2 & 12.39 &
$2\farcm0$ & $7\fdg2$  & 893 & 19.3 & 8724 & 0.00 & 0.84 \\
1760 & 4586 & & $\rm 12^h38^m28\fs1$ & $\rm  04^\circ19'09''$ & 1 & 12.06 &
$3\farcm9$ & $8\fdg3$  & 639 & 14.4 & 4657 & 0.00 & 0.98 \\
1859 & 4606 & & $\rm 12^h40^m57\fs7$ & $\rm 11^\circ54'46''$ & 1 & 12.28 &
$2\farcm9$ & $2\fdg5$ & 1528 & 12.7 & 7648 & 0.95 & 1.00 \\
\hline
\end{tabular}
\end{center}
$^a$ \object{NGC 4569} is also \object{M90}. 
\end{table*}

\subsection{Are the outlying \HI-deficient galaxies truly too far from the
Virgo cluster to have had their gas ram pressure stripped?} 
\label{wrongdist}
If one wishes to explain
the outlying \HI-deficient galaxies as galaxies that
have lost their interstellar 
gas by its stripping by
ram pressure from the hot intracluster gas, then these galaxies must 
have passed through the core of the Virgo cluster where the intracluster gas
is dense and the galaxy velocities large so that the ram pressure is largest.

\cite{MSSS03}
found
that objects that have passed through the
core of a structure in the past cannot be at distances greater than 1--2.5
virial radii from the cluster today. With their estimate of the distance (16.8
Mpc) and the virial radius (1.65 Mpc)
of Virgo, this means that galaxies that have passed through the core of Virgo
cannot lie at a distance from the Local Group greater than 18.5 or 20.9 Mpc (for $r_{\rm
reb}/r_{100} = 1$ or 2.5, respectively)
 nor can they lie closer than 15.1 or 12.7 Mpc from the Local Group.

For the 3 \HI-deficient spirals at 28 Mpc to have passed through the core of
Virgo, one would require that their distances be each overestimated by 52\%
(34\%), which  corresponds to $2.3\,\sigma$ ($1.6\,\sigma$) events for their
distance moduli, for $r_\mathrm{reb}/r_{100} = 1$ (2.5), respectively. Such
errors are possible, as shown in Sect.~\ref{role}.  Note that none of the
distances of the 13 galaxies listed in Table~7 are based upon Cepheid
measurements, which are much more precise than Tully-Fisher estimates. 
On the
other hand, given the presence in our sample of some spirals with grossly
incorrect distances (see Sect.~\ref{role}), one may wonder if the distances to
the 3 spirals at 28 Mpc could also be grossly incorrect: they could lie at the
cluster distance of 16.8 Mpc, thus leading in this case to distances
overestimated by 66\%, which for 20\% rms relative errors correspond to three
$2.8\,\sigma$ events for their distance moduli, which appears unlikely,
unless the Tully-Fisher error distribution has non-gaussian wings.

Similarly, the three foreground
\HI-deficient spirals at $\sim 10$ Mpc could have bounced out of the core of
the Virgo cluster, as they lie on the outer edge of the envelope of the
particles after inclusion of 20\% rms distance errors in Figure~\ref{errors}.
Indeed, were they bouncing out of the Virgo cluster in the foreground at 15.1
(12.7) Mpc, one would then have three cases of 34\% (21\%) errors, each
corresponding to $-2.3\,\sigma$ ($-1.3\,\sigma$) events for their distance
moduli, for $r_\mathrm{reb}/r_{100} = 1$ (2.5), respectively. Alternatively,
they could lie in the cluster proper, at 16.8 Mpc, which would correspond to
three $-2.8\,\sigma$ events for their distance moduli, which again appears
unlikely, unless the Tully-Fisher error distribution has non-gaussian wings.

The offset of a galaxy relative to the Tolman-Bondi locus in
Figure~\ref{proj} can be caused by a large distance error or a large peculiar
velocity relative to the general flow at that distance, or a combination of
both.
To see which effect is more important, one can consider the $N$ simulated
particles with distance (with errors folded in) and velocity respectively 
within 3 Mpc and $200 \, \rm km \,
s^{-1}$ of each galaxy in Table 7, as well as angular distance to the halo
center within $1\fdg5$ of that of the galaxy relative to M87, 
and
asked what fractions $P_1$ and $P_{2.5}$ of these particles were located
(before the distance errors were folded in)
within 1 or 2.5 times the virial radius of the halo, respectively.

These fractions must be taken with caution, because the
groups falling in or moving out of the Virgo cluster are 
at different distances and with different relative masses between the
simulations and the observations.
For example, a true group of galaxies in the background of the Virgo cluster
will end up with too high values of $P_1$ and $P_{2.5}$ if there is no group
at the same distance in the simulations, while a Virgo cluster galaxy with
a measured distance of 10 Mpc or so 
beyond the cluster will have too low values of
$P_1$ and $P_{2.5}$ if the simulations include a group at a distance close to
this measured distance.
Nevertheless, $P_1$ and $P_{2.5}$ should provide interesting first order
constraints. 

Of course, this method of locating galaxies, implies that galaxies
\object{NGC4064}, \object{NGC 4580} and \object{NGC 4586}, at angular
distances from M87 beyond $\theta_{100} + 1\fdg5 = 7\fdg1$ have $P_1
= 0$.
Interestingly, all three galaxies are likely to be within $2.5\,r_{100}$
(i.e. $P_2$ is close to unity).

Among the 5 foreground galaxies with $\theta <
5\fdg6$,
\object{NGC 4312}, \object{NGC 4438}, 
\object{M90} 
and \object{NGC 4606}
are all likely to be in the Virgo cluster 
and all very likely to be close enough that they may have passed through its
core.
Only \object{NGC 4424}, apparently at distance 4.1 Mpc, is too close to
actually be located within 2.5 virial radii from the Virgo cluster.

Among the 5 background  galaxies with $\theta < 5\fdg6$, all but
\object{NGC 4305}, lying at 40 Mpc, have a non negligible probability of having
crossed the cluster. Only \object{VCC 1569} measured at 23 Mpc is highly likely
to be located within the virial radius of  the Virgo cluster.
Interestingly, the 3 galaxies at measured distances around 28 Mpc have very
different values of $P_1$ and $P_{2.5}$.
The closest one, \object{NGC 4307}, has one chance in two of being located
near the cluster and 40\% probability of being within the cluster itself.
However, the furthest one, \object{NGC 4356}, has one chance in 5 of belonging
to the Virgo cluster and less than one chance in four of being within 2.5
virial radii.
The third one, \object{NGC 4492}, has only one chance in 9 of being within 2.5
virial radii of the Virgo cluster.
This low probability is not surprising, given that it is
located close to the infall/expansion zone of Figure~\ref{proj}.

\subsection{Are the outlying \HI-deficient galaxies located within small
groups?}
\label{envt}

The alternative to ram pressure stripping of gas is removal of gas
by tidal effects. One can envision various scenarios:
1) tidal stripping of gas beyond the optical radius;
2) tidal compression of gas clouds leading to gas transformation into stars;
3) tidal stripping of the gas reservoir infalling into the disk.
Such tides can occur from interacting galaxies or from the cluster or group
itself. We begin by asking whether the \HI-deficient galaxies apparently
outside of Virgo lie in groups.

\begin{figure*}
   \centering
\resizebox{!}{\vsize}{\includegraphics{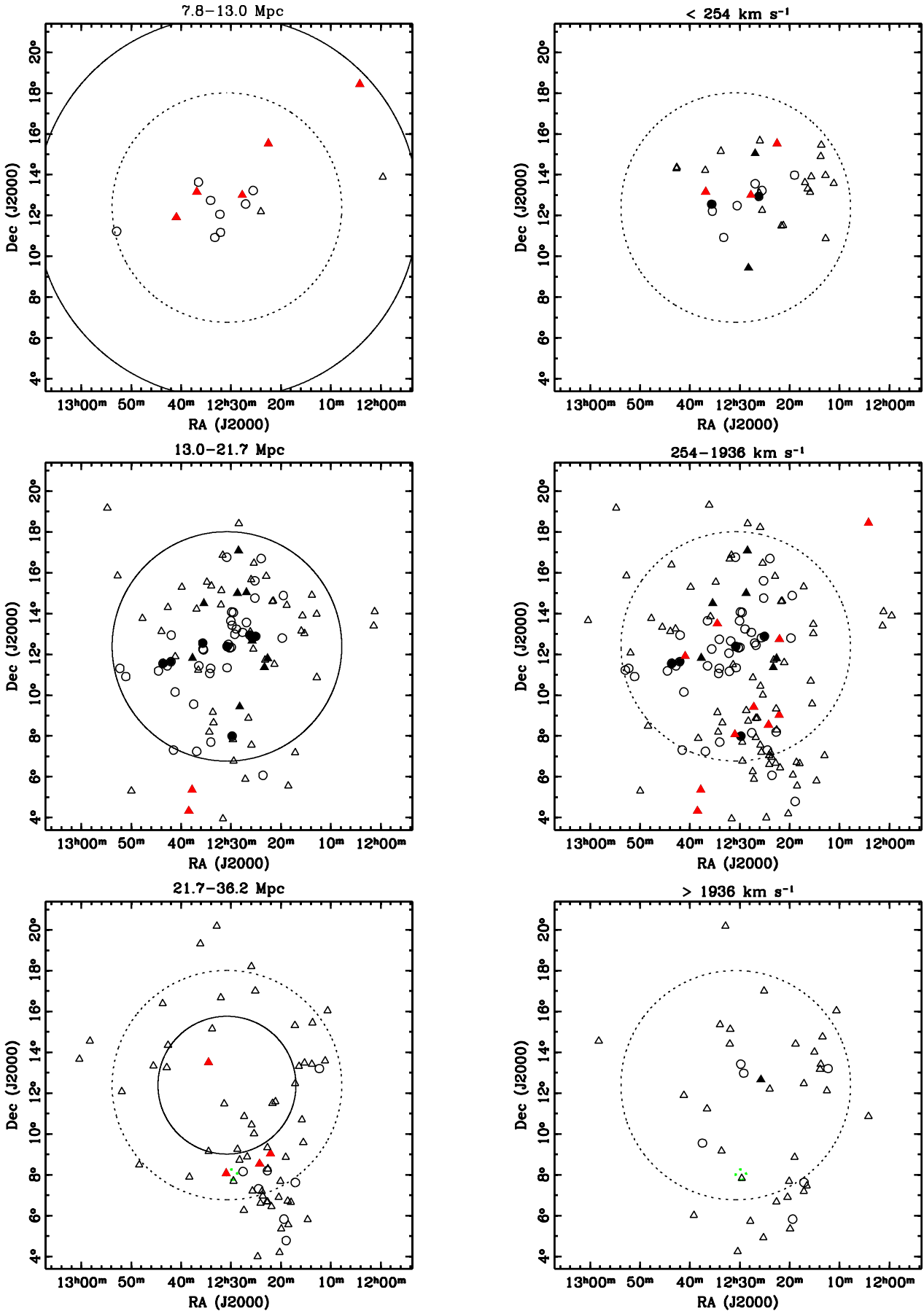}}
      \caption{Positions of galaxies within $9^\circ$ of M87 
in 3 distance (\emph{left}) and velocity 
   (\emph{right}) bins. Ellipticals and
   spirals are shown as \emph{circles} and \emph{triangles},
   respectively. Messier ellipticals are shown as \emph{filled circles},
   while \object{M49} (\object{NGC 4472}) is also highlighted as a
   \emph{thick dotted 
   circle} 
   (bottom plots, in \emph{green} in the Electronic version of the journal). 
\HIt-deficient ($>3\,\sigma$) spirals are shown as
   \emph{filled triangles} (with the 13 apparently in the outskirts of Virgo
   [Table 7] in \emph{red} in the Electronic version of the
   Journal). The \emph{large solid circles} show the virial 
   radius $r_{100}$ 
   at the    different    distance bins, while the \emph{large dotted
   circles} indicates $r_{100}$ at the distance to \object{M87}.}
          \label{windows5}
   \end{figure*}

Figure~\ref{windows5} shows the environment of the \HI-deficient galaxies in
three equal logarithm distance bins, $\pm 1.28\times\sigma \ln D$ wide,
centered around 10 Mpc (the apparent distance to the foreground \HI-deficient
galaxies), 16.8 Mpc (the distance to the Virgo cluster) and 28 Mpc (the
apparent distance to the background \HI-deficient galaxies), as well as in
two radial velocity bins. 
The log distance (velocity) 
bins are chosen wide enough that if the parent distribution
function of the log distance (velocity) errors were gaussian, one would have a
probability $2\,\mathrm{erf}(1.28)-1 = 0.86$ of having a galaxy with a true
distance (velocity) within the
interval appearing to lie at the center of the interval.

Among the 5 foreground \HI-deficient spirals, 4
(\object{NGC 4312}, \object{NGC 4438}, and \object{M90}, which, as noted in
Sect.~\ref{role}, is far off the Tolman-Bondi relation, and \object{NGC 4606}),
appear superposed (within
dotted circle in upper-left
plot) with the galaxies at the distance of the main structure of the Virgo
cluster, which suggests that
their distances may be seriously underestimated, unless their happens to be a
group of galaxies precisely along the line of sight to the center of Virgo
lying at 10 Mpc.

One may be tempted to identify the 3 \HI-deficient spirals close to
\object{M87} (\object{NGC 4438}, \object{M90}, and \object{NGC 4606})
with a group of ellipticals seen in the upper-left plot of
Figure~\ref{windows5}. The velocity dispersion of this group would explain
why these 3 galaxies are so far off the Tolman-Bondi solution given in
Figure~\ref{proj}. However, none of the ellipticals in this region have
accurate ($<38\%$)
relative distance estimates (cf. Table~2) 
and some or all may actually belong to the Virgo
cluster.

Interestingly, 3 of the 4 \HI-deficient spirals at 28 Mpc (\object{NGC 4307},
\object{NGC 4356} and \object{NGC 4492}) lie in a dense region, extending to
the SSW of the position of \object{M87} (lower-left plot), which is also seen
in the middle distance bin (middle-left plot) corresponding to the distance
to \object{M87}, but with 3 times fewer galaxies.  Although the eastern edge
of the region, and particularly the \HI-deficient \object{NGC 4492}, coincide
with the position of \object{M49} (\object{NGC 
4472}), it appears distinct from \object{M49}, which roughly lies at the
middle of a group of galaxies (middle-left plot) that does not seem to be
associated with the dense region at 28 Mpc (lower-left plot).

We now compare the Tully-Fisher distance estimator with the redshift distance
estimator $D_v = v/H_0$. 
If a structure outside of the Virgo cluster lies at a distance $D$
within a region with 1D velocity
dispersion $\sigma_v$, the error on the distance estimator $D_v = v/H_0$ will
be set by the dispersion of the peculiar velocities:
\begin{equation}
\delta D_v = {\sigma_v/H_0} \ .
\end{equation}
The redshift distance estimator $D_v$ will be more accurate than the direct
distance estimator when
\begin{equation}
H_0\,D > {\sigma_v \over \delta D/D} \ ,
\label{Dmin}
\end{equation}
where $\delta D/D$ is the relative error of the direct distance estimator.
For Tully-Fisher distance measurements, $\delta D/D \simeq 20\%$ and a field
velocity dispersion of $40-85 \, \rm km \, s^{-1}$ (as observed by
\citealp{Ekholm+01} and 
\citealp{Karachentsev+03}, as well as predicted by \citealp{KHKG03}),
equation~(\ref{Dmin}) 
yields a maximum distance for believing individual direct distances of 3 to
$7\,h_{2/3}^{-1}\,\mathrm{Mpc}$, so that individual galaxy distances of 28
Mpc are much less reliable.

This change in optimal distance estimators is illustrated in
Figure~\ref{figwedge}, where only one of the background \HI-deficient
galaxies has a radial velocity much larger than the mean of the Virgo
cluster.  The two plots on the right of Figure~\ref{windows5} show that the
structure to the South-Southwest (SSW) of the Virgo cluster lies close to
Virgo in velocity space (i.e. the concentration is in the right middle plot).
In redshift space, 
the 3 background \HI-deficient galaxies to the SSW of \object{M87} remain in
a similar concentration of galaxies (middle right plot of
Fig.~\ref{windows5})
as in distance space (lower left plot), but this
concentration is at similar radial velocities as the Virgo cluster.
However, only one
(\object{NGC 4307}) has a reasonably high
chance ($P_1$ in Table 7) of being located in or near the Virgo cluster,
while one (\object{NGC 4356}) has less than one chance in 4 of being in or near
the cluster (from the analysis of Sect.~\ref{wrongdist}), 
and one (\object{NGC 4492}) has a radial velocity indicative of a background
object 
expanding away from the Virgo cluster (this galaxy shows up as the empty 
triangle 
in Fig.~\ref{proj} very close to the Tolman-Bondi locus), and 
has only roughly one chance in 9 of being even close to the cluster.  
But the analysis of $P_1$ and $P_{2.5}$ of Sect.~\ref{wrongdist} should be
superior to the analysis of Fig.~\ref{windows5}, 
because it takes into account the correlation of distance and
velocity.
In summary, the 3 background \HI-deficient galaxies to the SSW of
\object{M87} may be suffering from tidal effects of a background group of
galaxies.

\subsection{Close companions and mergers}
\label{compmerg}
According to a search in {\sf SIMBAD}, we find that (at least)
4 of the 13 \HI-deficient galaxies in 
Table~7
have companions whose mass and proximity may generate tides that could remove
some of the neutral Hydrogen, either by stripping the gas beyond the optical
radius (without stripping the stars, thus leaving an \HI-deficiency) or by
compressing the diffuse neutral hydrogen, converting atomic gas into
molecular gas, then into stars. 
Of course, the presence of a companion in projection, even
with a similar 
velocity, does not guarantee that this companion is in physical interaction
with the galaxy in question, as projection effects are important in clusters.

\begin{table}[ht]
\label{companions}
\caption{Companions to \HIt-deficient galaxies in the outskirts of Virgo}
\tabcolsep 3pt
\begin{center}
\begin{tabular}{llcllccccc}
\hline
\hline
\multicolumn{2}{c}{Galaxy} & 
&
\multicolumn{2}{c}{Companion} & 
\multicolumn{1}{c}{$\Delta v$} &
\multicolumn{1}{c}{$\Delta m_T^c$} &
\multicolumn{1}{c}{$d/r_\mathrm{opt}$} & 
M & K \\
\cline{1-2}
\cline{4-5}
\multicolumn{1}{c}{VCC} & \multicolumn{1}{c}{NGC} & & \multicolumn{1}{c}{VCC} &
\multicolumn{1}{c}{NGC}  & ($\rm km \, s^{-1}$) & & & \\
\hline
\ \,522 & 4305     & &  \ \,523 & 4306 & $-380$ & \ \,0.9$^a$ & 2.8 & R? &
--- \\
1043    & 4438$^b$ & & 1030  & 4435 & $-214$ & 1.0 & 1.0 & I & I \\
1690    & 4569$^c$ & & 1686$^d$ & \multicolumn{1}{c}{---}   & \ 1354       &
4.2 & 1.1 & R & R \\ 
1859    & 4606     & & 1868 & 4607 & \ \ \,593   & 0.6  & 2.7 & I? & --- \\
\hline

\end{tabular}
\end{center}
\noindent Notes:
$^a$ $B$-band;
$^b$ also Arp 120;
$^c$ also M90 and Arp 76;
$^d$ also IC 3583.
\end{table}

Table~8 shows the properties of the companions to these 4 galaxies.
Column 5 is the difference in velocity relative to the Local Group; 
column 6 is the difference in extinction-corrected total $I$-band magnitude;
column 7 is the angular distance between the two components, in units of the
optical radius ($D_{25}/2$) of the major member;
columns 8 and 9 are indicators for morphological and kinematical
\citep{RWK99} disturbances (R $\to$ regular, I $\to$ irregular) for the major
component.
Table 8 indicates that \object{NGC 4438} has a very close major companion,
which is likely to be responsible for its irregular morphology and internal
kinematics.
\object{NGC 4305} and \object{NGC 4606} have moderately close major
companions.
\object{M90} has a close very minor ($\approx 1/50$ in mass) companion.

Moreover, \object{NGC 4307} and \object{NGC 4356} are close in 4D space: their
projected separation (at the distance of M87) is 215 kpc and their velocity
difference is $85 \, \rm km \, s^{-1}$, so they may constitute an interacting
pair. 
Similarly, \object{NGC 4580} and \object{NGC 4586} are at a projected
separation of 311 kpc with a velocity difference of $254 \, \rm km \,
s^{-1}$, so they too may constitute an interacting pair.

Finally, two other galaxies in Table 7 have unusually low inner rotation
\citep{RWK99}: 
\object{NGC 4064} and
\object{NGC 4424}.
\cite{KKRY96} argue that \object{NGC 4424} has undergone a merger with a
galaxy 2 to 10 times less massive  
and presumably the same
can be said for \object{NGC 4064}.

We can obtain some clues about the origin of the \HI\ deficiency of the
outlying galaxies by studying their star formation rates. 
We adopt the,
obviously simplified, point of view  that tidal
interactions are believed to enhance star formation
(e.g. \citealp{LK95}),
while ram pressure stripping should reduce star formation.
We therefore verify the tidal interaction hypothesis by checking for enhanced
star formation rates.
Because tidal interactions operate on scales of order the orbital times of
interacting pairs, typically 250 Myr or more, we do not use the H$\alpha$
star formation rate indicator, as it is only sensitive to very recent star
formation, but focus instead on the broad-band color $B\!-\!H$.
We have taken $B$ and $H$ magnitudes from the {\sf GOLDMine} database
\citep{Gavazzi+03}.

\begin{figure*}[ht]
\centering
\resizebox{0.9\hsize}{!}{\includegraphics{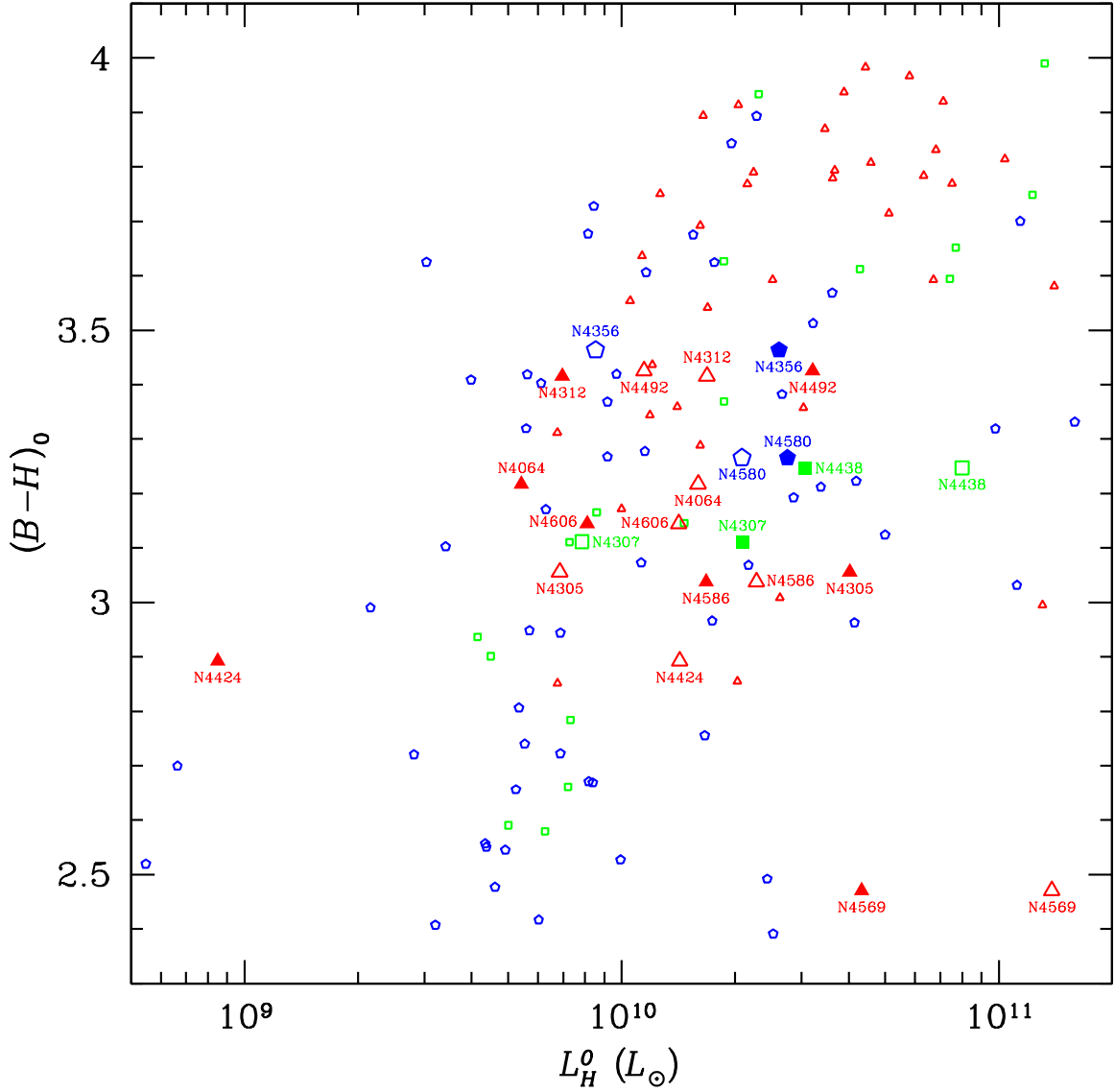}}
\caption{Color-luminosity diagram for
the 12 of the 13 \HIt-deficient galaxies of Table 7 (\object{VCC~1569} was
excluded because it is not bright enough to be included in {\sf GOLDMine}),
possibly outside the Virgo cluster  
(\emph{large filled symbols}) or placed at the
distance of the Virgo cluster (\emph{large open symbols}).
Other Virgo
galaxies are shown as \emph{small open symbols}.
\emph{Triangles}, \emph{squares} and \emph{pentagons} (respectively 
in \emph{red},
\emph{green}, and \emph{blue} in the Electronic version of the journal)
refer to galaxies of
morphological type $1-2$, $3-4$ and $5-6$, respectively.
The data is taken from the {\sf GOLDMine} \citep{Gavazzi+03} database, and
dereddended using the internal and Galactic extinctions from {\sf LEDA}
with the selective extinctions of \cite*{CCM89}.
}
\label{BmH}
\end{figure*}

Figure~\ref{BmH} shows the $B\!-\!H$ 
colors as a function of $H$-band luminosity
for the 12 NGC galaxies of Table~7 (we have omitted \object{VCC~1569}, which is
too underluminous to be in {\sf GOLDMine})
and for other Virgo galaxies. The colors and
luminosities in this figure were corrected for absorption using internal and
galactic \citep*{SFD98} absorption obtained from {\sf LEDA}, assuming
selective extinction coefficients from \cite*{CCM89}.
Figure~\ref{BmH} clearly displays a general color-luminosity relation,
i.e. the trend that more luminous galaxies are redder.
Comparing galaxies of similar mophological types, 
one checks that among the four candidates for
possessing tidal companions, 
\object{NGC~4305},
\object{NGC~4438},
\object{NGC~4569} (\object{M90}),
and 
\object{NGC~4606} (see Table 8), the \emph{first} three are 
blue for their luminosities, assuming their Tully-Fisher distances, 
but given that the \emph{latter} three might well
lie in Virgo (see column $P_1$ in Table 7), we also check that at the 
distance to the Virgo cluster 
these 3 galaxies are very blue (\object{NGC~4438}), extremely blue
(\object{M90}) and just blue (\object{NGC~4606}). The blue colors for these
4 galaxies, given their luminosity and morphological type, confirms the tidal
hypothesis for all four objects.

Interestingly, the possibly interacting galaxies \object{NGC~4307} and
\object{NGC~4356} have blue colors confirming their interaction if they are
indeed in the background of Virgo as their Tully-Fisher distances suggest,
but have red colors as expected for ram pressure stripped galaxies if they
lie in the 
Virgo cluster. The values of $P_1$ in Table~7 cannot distinguish between these
two possibilities.
The situation is even more uncertain for the possibly interacting galaxies 
\object{NGC~4580} and \object{NGC~4586}, whose Tully-Fisher distances place
them slightly behind and in front of the Virgo cluster, respectively.
If they both lie at the distance of the Virgo cluster, the first one would
have a normal color, while \object{NGC~4586} would be blue as expected if it
is indeed interacting with \object{NGC~4580}.

The situation is also confusing for the two galaxies with possible recent
minor  merging: \object{NGC~4064} appears normal or red if it lies in the
foreground as suggested by its Tully-Fisher distance, but blue if it lies at
the distance of the Virgo cluster, and \object{NGC~4424} appears normal or
red if it lies as close as its Tully-Fisher distance suggests but quite blue
if it lies at the distance of Virgo.
One would expect the minor merger to generate a blue nucleus if the merging
took place recently enough, but this may have little effect on the overall
color of the galaxy.

Moreover, 
one may contest that tides will remove stars as well as gas from the disk of
a spiral galaxy.
However, 
the estimates of \HI-deficiency presented here are based upon single-dish (Arecibo) radio
observations, which often do not have sufficient angular resolution to map the gas
(the half-power beam of Arecibo subtends $3\farcm3$, close to  
the typical optical angular diameters of the \HI-deficient galaxies --- see
Table~7). 
One can therefore easily imagine that fairly weak tidal effects will deplete
the neutral Hydrogen gas beyond the optical radius without affecting the stars in the
spiral disk, thus leading to an \HI-deficiency when normalized relative to
the size of the optical disk.

Two galaxies among our \HI-deficient Virgo outliers listed in 
Table~7
(\object{NGC 4438} and \object{M90}) have been studied by \cite{CKBvG94},
who computed \HI-deficiencies of Virgo spirals, based upon high resolution 21cm
VLA observations. Both galaxies present very small 21cm to optical diameter
ratios, indicating that the gas deficiency sets in at small radii and cannot be
caused by tidal effects. Additional high-resolution 21 cm observations 
at the VLA 
of the other galaxies in Table~7 
will obviously allow confirmation of this result.

\subsection{Are the outlying \HI-deficient galaxies truly
gas-deficient?}

\label{trulydef}

The \HI-deficiency estimator has been defined by several authors 
(e.g. \citealp{CBG80}; \citealp{HG84}; \citealp{SGH96}) as
the hydrogen mass relative to the
mean for galaxies of the same morphological type and optical
diameter:
\begin{equation}\label{defusual}
\hbox{DEF}_1  =
\langle\log\mhi(D_\mathrm{opt},T)\rangle-\log\mhi\;,
\end{equation} 
\noindent
where $\mhi$ is the \hi\ mass of the galaxy in solar units.
\cite{Solanes+02} applied a different
\HI-deficiency estimator based on the hybrid mean surface brightness expected
for each galaxy type independently of its diameter,
which has the advantage of being independent of the distance to the galaxies:

\begin{equation}\label{defsolanes} 
\df_2 = \langle\log \shi (T)\rangle-\log\shi\ ,
\end{equation}
where $\shi$ is the hybrid mean surface brightness, defined as
\begin{equation}
\shi = {F_\mathrm{H{\tiny I}} \over \theta_\mathrm{opt}^2} \ ,
\end{equation}
where $F_\mathrm{H{\tiny I}}$ and $\theta_\mathrm{opt}$ are the \HI\ flux and
optical angular diameter, respectively.

\begin{figure}[ht]
\centering
\resizebox{9cm}{!}{  \includegraphics{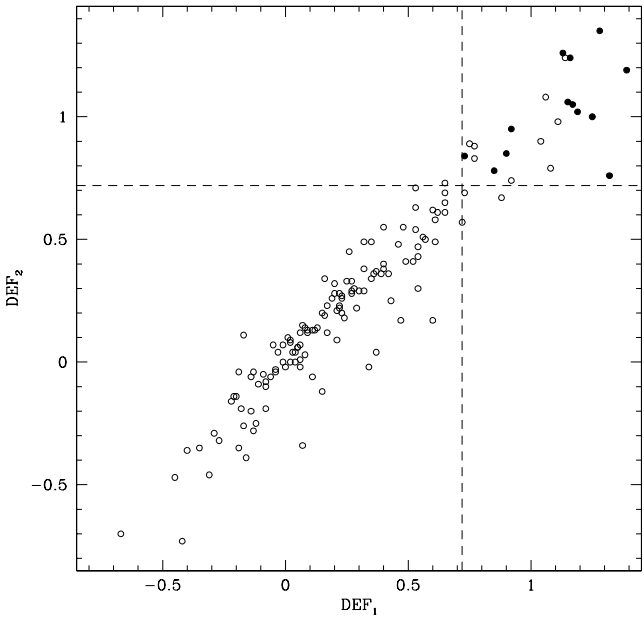}}
\caption{Comparison of \HIt-deficiency estimators.
Galaxies possibly located outside the Virgo cluster 
are shown as \emph{filled circles}.
The \emph{vertical} and \emph{horizontal dashed lines} shows the $3\,\sigma$
cutoff for DEF$_1$ and DEF$_2$, respectively.
}
\label{def1def2}
\end{figure}

Could the seemingly \HI-deficient galaxies in the foreground and background
of Virgo have normal gas content according to equation~(\ref{defusual}) but
appear deficient with equation~(\ref{defsolanes})?
\citeauthor{Solanes+02} checked the statistical agreement of these two
\HI-deficiency estimators by fitting a linear model between the two
parameters and found a regression line with slope and intercept close to the
unit and zero respectively, with a scatter or 0.126. 
Figure~\ref{def1def2} shows this correlation, where those galaxies possible
located outside the Virgo cluster are
shown as solid circles. The plot clearly indicates that the outliers are also
deficient with the $\df_1$ definition of \HI-deficiency.

However, 
Table~7
clearly shows that 7 out of our 13 \HI-deficient possible
outliers appear to be Sa galaxies ($T=1$), some of which might be misclassified
S0/a or even S0 galaxies, which are known to have little cold gas.
The stronger \HI-deficiency of early-type spirals in the Virgo cluster 
has been already noticed by
\cite*{GRV85}.
Some of 
these Sa galaxies may not be deficient relative to field S0/a or S0 galaxies.
Indeed, the most recent study \citep*{BGG03} of the \HI\ content of galaxies as
a function of morphological type indicates that the logarithm of the
\HI\ mass normalized to
luminosity or to square optical diameter decreases increasingly faster for
earlier galaxy types, therefore its slope for Sa's is considerably larger
than for Sc's, and hence S0/a galaxies misclassified as Sa's will appear more
\HI-deficient than later-type galaxies misclassified by one-half of a  
morphological
type within the Hubble sequence.

Moreover, independent estimates of morphological types by experts led to
$\sigma (T) = 1.5$ (rms) for galaxies with optical diameters $> 2'$
\citep{Naim+95}. The galaxies listed in 
Table~7
have optical diameters ranging
from $2'$ to $10'$ plus one (\object{VCC 1569}) with $D = 0\farcm8$. There
are two 
galaxies (\object{NGC 4305} and \object{NGC 4492})   with $T = 1$ and $D \leq
2'$, i.e. where the uncertainty on $T$ is probably as high as 1.5. 

For example, \object{NGC 4424} and \object{NGC 4492} are marginally 
\HI-deficient and are prime candidates for losing their status of gas-deficient
if they are misclassified lenticulars (by $\Delta T = 1$, the typical scatter
given by {\sf LEDA}). 

\begin{figure*}[ht]
\centering
\resizebox{\hsize}{!}{\includegraphics{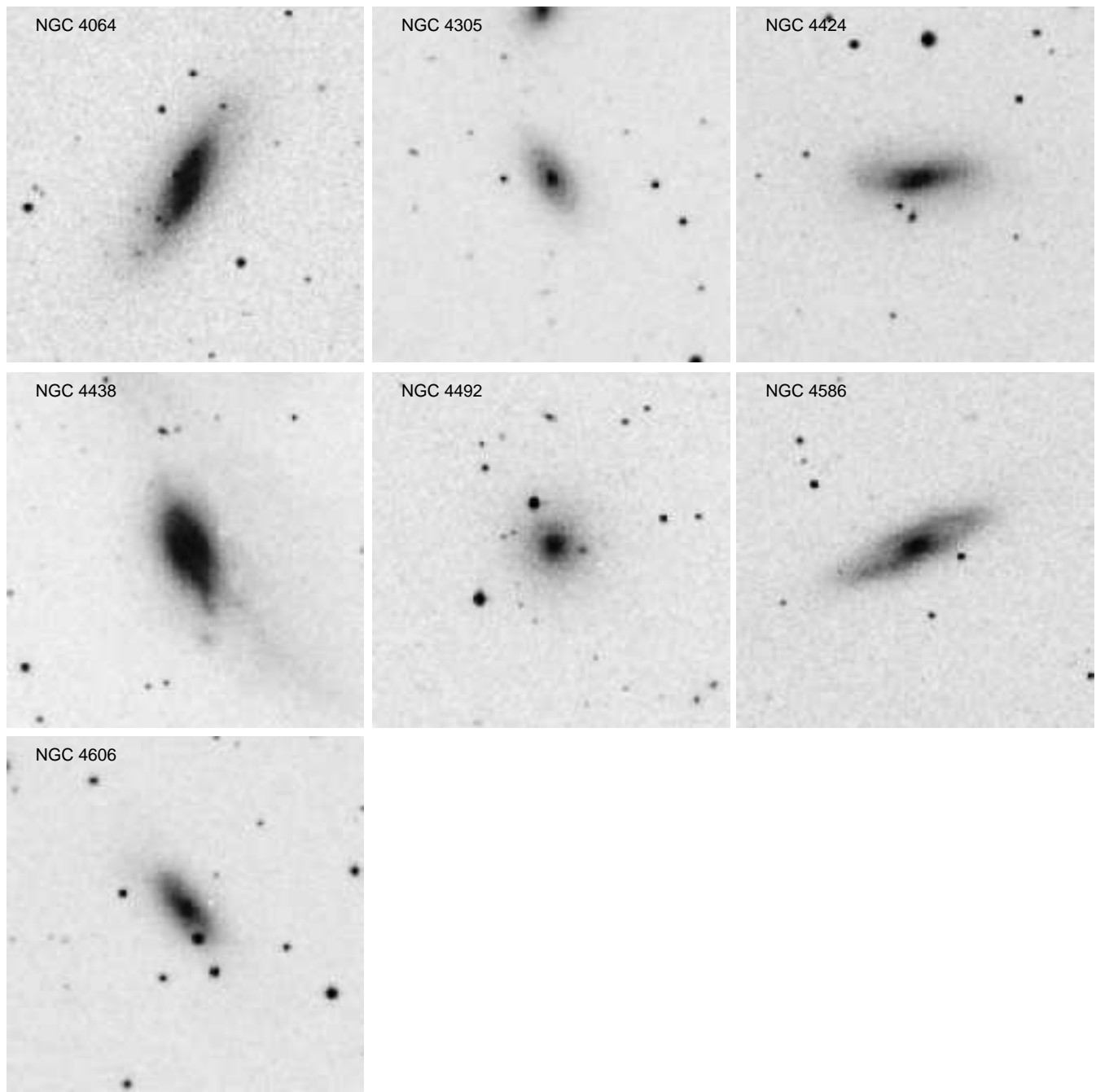}}
\caption{POSS I snapshots of the $T=1$ \HIt-deficient galaxies of Table~7. The
box size is $6'$ and North points upwards.}
\label{mosaic}
\end{figure*}

We must thus resort to visualizing the possibly misclassified galaxies.
Figure~\ref{mosaic} displays the snapshots of the 7 \HI-deficient 
outliers classified as Sa ($T=1)$.
We now discuss the morphologies of these 7 galaxies.
\begin{description}
\item [{\bf \object{NGC 4064}}] This galaxy 
has a very large high surface brightness inner disk or flattened bulge 
with little trace of spiral arms. 
\object{NGC 4064} may lie between S0 and S0/a.
\item [{\bf \object{NGC 4305}}] This galaxy is of type Sa or later. It has a
nearby major companion.
\item [{\bf \object{NGC 4424}}] This galaxy has an important bar and the size
of the bulge is difficult to 
estimate (\citealp{KKRY96} fit the bulge-disk ratio to between 0.2 and 0.6).
It shows little traces of spiral arms.
\citeauthor{KKRY96} write ``\object{NGC 4424} is
classified as an Sa rather 
than an S0 because the outer stellar disk is not structureless''.
It may well be an S0/a.
\item [{\bf \object{NGC 4438}}] This galaxy, with a large bulge, 
is tidally distorted by a companion
just North 
of the frame. \cite{Gavazzi+00} fit a pure $r^{1/4}$ law to this galaxy,
suggesting that its morphology is of an earlier type than Sa.
\item [{\bf \object{NGC 4492}}] This galaxy is viewed face-on, hence the disk
is not very 
visible. The bulge is relatively large and their are signs of a weak 
spiral arm.
\object{NGC 4492} may lie between S0/a and Sa, but \cite{Gavazzi+00} fit a
bulge to disk ratio of 1/3, suggesting a type of Sa or later.
\item [{\bf \object{NGC 4586}}] This galaxy shows spiral arms and a normal
bulge for an Sa galaxy.
\item [{\bf \object{NGC 4606}}] This galaxy shows little traces of spiral
arms and has an 
intermediate bulge. \object{NGC 4606} appears between S0/a and Sa.
However, \cite{Gavazzi+00} fit a bulge to disk ratio of 1/20, suggesting a
later type than Sa.
\end{description}
Altogether, we
suspect that 2 galaxies may be earlier than Sa: \object{NGC 4064} and
\object{NGC 4424}, both of which  
appear to have recently ingested smaller galaxies, and
significant amounts of cold gas may have been heated up during the merger.

Given that between types S0/a and Sa, the compilation of \cite{BGG03} (left
plots of their Fig.~2) indicates that the normalized \HI\ content varies as
$\log M_\mathrm{HI}^\mathrm{norm} \sim T/4 + \mathrm{cst}$, an overestimate
of 1 (0.5) in 
$T$ will lead to an underestimate in \HI-deficiency of 0.25 (0.125). 
The \HI-deficiency of \object{NGC 4064} is sufficiently high that a
correction of 0.25 will still keep it among the $3\,\sigma$ cases using both
DEF$_1$ and DEF$_2$ criteria.
On the other hand, if the morphological type of \object{NGC 4424} 
is S0/a ($T=0$), it would still be \HI-deficient
with criterion DEF$_1$, but no longer among the $3\,\sigma$ cases with
DEF$_2$.

Note, however, that if most \HI-deficient outliers classified as Sa's were in
fact S0/a's, we would find
ourselves with the new problem of justifying the presence of S0/a's so far away
from the cluster core and apparently not within groups (see section 5.3).
We also have to keep in mind that \cite{KK98} argued that spirals in the Virgo
cluster could have also been misidentified as early-type objects, an opposite
trend to the above suggested.

\subsection{Case by case analysis}

We now focus on each of the 13 galaxies, one by one.

\begin{description}
\item [{\bf \object{NGC 4064}}] The 
very low inner rotation \citep*{RWK99} suggests that it is a
merger remnant (see \object{NGC 4424} below), which may have been responsible
for its \HI-deficiency. 

\item [{\bf \object{NGC 4305}}] This galaxy, at 40 Mpc and with a large
radial velocity, is almost certainly a background galaxy. It 
has a major tidal companion (Table~8) and a blue color for its luminosity,
assuming it is indeed
far behind Virgo, confirming the tidal origin of its gas deficiency.

\item [{\bf \object{NGC 4307}}] The origin of the \HI-deficiency in this
galaxy is difficult to ascertain, as it has 41\% (48\%) probability of
lying within the sphere centered on the Virgo cluster with radius its 1 (2.5) 
times the virial radius.
Its proximity with \object{NGC 4356} suggests that it may have interacted
in the past with that galaxy, and if it is indeed in the background of Virgo,
as suggested by its Tully-Fisher distance, it would be blue for its
luminosity, confirming the interaction hypothesis with \object{NGC~4356}.

\item [{\bf \object{NGC 4312}}] This galaxy is very likely within the Virgo
cluster.

\item [{\bf \object{NGC 4356}}] The origin of the \HI-deficiency in this
galaxy is difficult to ascertain, as it has 19\% (23\%) probability of
lying within the sphere centered on the Virgo cluster with radius its 1 (2.5) 
times the virial radius.
Its proximity with \object{NGC 4307} suggests that it may have interacted
in the past with that galaxy, as confirmed by its blue color for its
luminosity, assuming it lies indeed in the background of Virgo.

\item [{\bf \object{NGC 4424}}] The very low inner rotation of this galaxy 
suggests a merger remnant
\citep{RWK99}.  
It almost certainly lies in the
foreground of the Virgo cluster.
Moreover, as discussed in Sect.~\ref{trulydef},
\object{NGC~4424} may be of type S0/a, in which case it would not be
$3\,\sigma$ deficient in neutral Hydrogen.

\item [{\bf \object{NGC 4438}}] 
This seemingly foreground galaxy is very probably a member of the Virgo
cluster.
\cite{KE83} suggest that the \HI-deficiency of this highly distorted galaxy
is caused not by 
the tidal companion, which should increase the line-width, but by interaction
with the intracluster medium of the Virgo cluster. Moreover, this galaxy has
a very small radial extent of its \HI\ gas, as attested by the VLA map of 
\cite{CKBvG94}.
Interestingly, at the distance of Virgo, this galaxy would be very blue,
which is easily explained by the presence of its tidal companion.

\item [{\bf \object{NGC 4492}}] This galaxy is probably in the background of
the Virgo cluster, in which case it would be blue for its luminosity and
morphological type.
It may lie in a group that could include \object{NGC~4307}
and \object{NGC~4356} and thus be suffering tidal encounters with these
and/or other members of the group.

\item [{\bf \object{VCC 1569}}] This faint galaxy probably lies in or very
close to the Virgo cluster.

\item[{\bf \object{M90}}]
Among the 13 galaxies listed in 
Table 7,
this very blue Sab LINER (e.g. \citealp{Keel96})
deviates the most significantly from both the Tolman-Bondi locus
(Fig.~\ref{proj}) and from the
color-luminosity diagram (Fig.~\ref{BmH}).
Its distance has been controversial.
\cite*{SKY86} suggest that the Tully-Fisher estimate is an underestimate
because the optical rotation curve is still rising to the limit of its
measurement (see \citealp{RWK99}) and that similarly low distances derived
by \cite*{CCM82} using their line index / bulge luminosity relation may be
erroneous because the weakness of the nuclear lines is in contrast with the
high metallicity expected from the stellar population of the bulge.

\object{M90} is extremely isolated in projected phase space.
Among the galaxies with $D < 13\,\rm Mpc$, it appears associated with a group
of ellipticals, none of which has accurate distance measurements, with
sky coordinates associated with the central ellipticals with accurate
distance measurements in the center of the Virgo cluster.
Given its bright corrected total blue magnitude of 9.63 (Table 7),
it's absolute magnitude would correspond to 
$M_B = -20.24$ at 9.4 Mpc or $M_B = -21.5$ at 16.8 Mpc.
The latter absolute magnitude is very bright for a spiral galaxy.
This can be quantified by comparing with the turnover luminosity, $L_*$,
of the field galaxy 
luminosity function. Unfortunately, the surveys used to derive the field
galaxy luminosity function do not employ the Johnson $B$ band used here.
However, the {\sf 2MASS} survey has measured a total $K$-band magnitude of $K
= 6.58$ for \object{M90}. For a turnover absolute magnitude \citep{Kochanek+01}
$M_K^* = -23.39 + 5\,\log h = -24.27$ for $h=2/3$ adopted here, the $K$-band
luminosity of \object{M90} corresponds to $0.4\,L_*$ if \object{M90} is at
9.4 Mpc but $1.3\, L_*$ if it lies at the distance of the Virgo cluster, which
is a large luminosity for a spiral galaxy.
Moreover, at the distance of Virgo, \object{M90} would be extremely blue for
its luminosity (Fig.~\ref{BmH}).

Finally, the contours of continuum radio emission seen from the {\sf NVSS}
survey taken at the VLA are compressed in the East-Northeast direction, which
is opposite to the direction towards the center of the Virgo cluster.
This suggests that \object{M90} may be bouncing out of the cluster, which
would also explain the galaxy's negative radial velocity.

\item [{\bf \object{NGC 4580}}] This galaxy probably lies near but not in
the Virgo
cluster, and may possibly be bouncing out of the cluster, therefore losing
its neutral gas by ram pressure stripping.
It may be interacting with \object{NGC 4586}, and is somewhat
blue for its luminosity, especially if it lies behind Virgo.

\item [{\bf \object{NGC 4586}}] This galaxy probably lies near the Virgo
cluster and may possibly be bouncing out of the cluster, therefore losing
its neutral gas by ram pressure stripping.
It may be interacting with \object{NGC 4580}, and is 
blue for its luminosity, especially if it lies within the Virgo cluster.

\item [{\bf \object{NGC 4606}}] This galaxy is a very likely member of 
the Virgo
cluster.
It is likely to be tidally perturbed
by its companion \object{NGC 4607} (Table~8), as attested by its blue color
for its luminosity assuming it indeed lies in Virgo,
although it shows only modest
signs of such tides.
\end{description}

\subsection{Origin of the \HI-deficiency in the outlying galaxies}

The \HI-deficient galaxies apparently lying well in front and behind
of the Virgo cluster must have passed through the core
of the cluster to have shed their gas by stripping by the ram pressure
of the intracluster gas. As shown by \cite{MSSS03}, galaxies cannot bounce
out beyond 1 to 2.5 virial radii, which according to their 
estimate of the virial
radius of Virgo, amounts to 1.7 to 4.1 Mpc. 
Therefore, given a distance to the Virgo cluster of 16.8 Mpc, we suggest that
only galaxies with distances between 12.7 and 20.9 Mpc could have passed
through Virgo.

One
possibility, discussed in Sect.~\ref{wrongdist}, is that the true 3D locations
of the outlying
\HI-deficient galaxies are in fact within the Virgo cluster's virial radius
or at most within 4 
Mpc from its center, which is possible (although not likely) given the typical
20\% relative distance errors for the spiral galaxies in Virgo, expected from
the Tully-Fisher relation, and from our comparison of the observed velocity
distance relation with that derived from cosmological $N$-body simulations.
Indeed, of the 13 \HI-deficient spirals appearing in the outskirts of the
Virgo cluster, 4 to 5 (depending on how to place the distance to
\object{M90}) are highly likely to be located in 3D within the virial radius
of the 
Virgo cluster (Table 7) and one other galaxy (\object{NGC 4307}) has one
chance in two of being in the cluster or near it. Three others are at an
angular radius $> \theta_{100}$ from M87, but appear to be within 2.5 virial
radii of it in 3D. 
Finally, two galaxies (\object{4356} and \object{NGC 4492}) 
in the direction of the cluster have low but non
negligible probability of being in or near the cluster, while the last two
are definitely foreground (\object{NGC 4424}) and background (\object{NGC
4305}) galaxies.
Hence, between 4 and 11 of these 13 galaxies may have passed through the core
of the Virgo cluster and seen their neutral gas ram pressure swept by the hot
intracluster gas.

The presence of a few discrepant distance measurements suggests that the
nearly gaussian parent distribution function for measurement errors in the
distance modulus (or relative distance) has extended non-gaussian wings.

It is also possible that highly inaccurate (at the 30\% level or more)
Tully-Fisher distance measurements may arise from
physical biases.
If galaxies are \HI-deficient because of strong interactions with
the cluster (through ram pressure stripping or tides) or with neighboring
galaxies (through collisional tides), their Tully-Fisher distances may be
biased and/or inefficient. 
Indeed, our analysis of Sect.~\ref{compmerg} indicates that,
among the galaxies likely or possibly within the Virgo cluster, 
one (\object{NGC 4438}) 
shows clear tidal perturbations from a close major companion, while 
2 others (\object{M90} and \object{NGC 4606}) are 
probably tidally perturbed by close companions, and the one possible Virgo
cluster member (\object{NGC 4307})
may have interacted
with a fairly close major companion in the past.

Moreover, there are general arguments, independent of these case-study tidal
effects, that suggest that
abnormal Tully-Fisher distances may be
caused by \HI-deficiency.
For example, \cite{SKY86} noted that \HI-deficient spirals in the direction
of the Virgo cluster tend to have
smaller line-widths for their $H$-band luminosity, and, similarly,
\cite*{RHF91} found that the maximum rotation velocity  of spiral
galaxies within compact groups of galaxies  
tend to be smaller for their luminosity than
for normal spirals.
Such small line-widths or maximum rotation velocities  
imply that distances are underestimated.
Using 2D (Fabry-Perot) spectroscopy,
\cite{MAPB03} have refined the analysis of \citeauthor{RHF91}
and find that the lowest luminosity galaxies in their sample ($M_B = -19.5\pm
1$)  have lower
maximum rotation velocities than field spirals of the same
luminosity. Moreover, they find that compact group galaxies display a more
scattered Tully-Fisher relation than that of field spirals.

For galaxies lying in or very near the Virgo cluster, 
an alternative to ram pressure gas stripping is for tidally stripped gaseous
reservoirs that prevent subsequent gas infall to the disk, which quenches
subsequent star formation \citep*{LTC80}. 
However, if a spiral galaxy lies in a cluster, it may well be on its first
infall, provided that the effective ram pressure stripping of its neutral gas
quickly converts it into an S0 galaxy.
It seems reasonable to expect that 
tidally stripped gas reservoirs will not have time to have a large effect on
the gas content, so that such galaxies with tidally stripped gas reservoirs
will not be as severely \HI-deficient as the 13 $3\,\sigma$ cases of Table~7.

The \HI-deficient galaxies that are truly outside of the cluster 
may have lost their gas by
the tides generated during close encounters with other galaxies within a
group or small cluster of galaxies.
Indeed, the efficiency of tidal perturbations of galaxies from close
encounters with other galaxies is stronger in groups and 
small clusters than in rich clusters (see Fig.~5 of \citealp{M00_IAP}),
basically because the rapid motions in rich clusters decrease the efficiency
of tides.
Three of the four \HI-deficient galaxies at 28 Mpc lie within a significant
concentration of galaxies (Figs.~\ref{figwedge} and \ref{windows5}) that
ought to create such collisional tides, even with
the estimates of distance using the
more precise radial velocities (for these distant objects, see
eq.~[\ref{Dmin}]).

The 3 galaxies outside the virial angular radius (\object{NGC 4064},
\object{NGC4580} and \object{NGC4586}) all probably lie
closer than 2.5 virial radii from M87, and may be in the process of bouncing
out of the cluster after having lost their \HI\ gas through ram pressure
stripping by the intracluster gas, if the maximum rebound radius is closer to
the upper limit of 2.5 virial radii found by \cite{MSSS03}, than of their
lower limit of 1 virial radius.
One of these galaxies
(\object{NGC 4064}) is a candidate for a recent merger as witnessed by its
very low inner rotation, while the other two (\object{NGC 4580} and
\object{NGC 4586}) may have interacted in the past, given their proximity in
phase space.
In the case of recent merging, the neutral gas may have been shock heated
with that of the galaxy being swallowed up, and should shine in H$\alpha$
or X-rays depending on its temperature.

Of the 4 galaxies likely or certainly far outside
the Virgo cluster (in 3D), one (\object{NGC 4305}) may be tidally perturbed
by a close major companion, one (\object{NGC 4424}) has suffered a recent
merger (given its very low inner rotation velocity), one may have interacted
with another galaxy (\object{NGC 4307}) in the past, and one (\object{NGC
4492}) may be a misclassified S0/a (with $T=0$), implying a lower expected
\HI\ content, hence a less than $3\,\sigma$ \HI-deficiency upon the estimator
DEF$_1$ and just at $3\,\sigma$ upon DEF$_2$, although the bulge/disk ratio
derived by \cite{Gavazzi+00} suggests that the galaxy is indeed an Sa.

Although we can provide at least one explanation 
for each case of \HI-deficiency in the 13 spiral
galaxies that 1) appear to be located in the foreground or background of the
Virgo cluster or 2) lie just outside the projected 
virial radius of the cluster, there is certainly room for an improved
analysis. In particular, accurate distances obtained with Cepheids would be
most beneficial, as well as VLA maps for 11 of the 13 galaxies to distinguish
tidal stripping from ram pressure stripping.

\begin{acknowledgements}
We wish to thank Philippe Amram, Hector Bravo-Alfaro,
V\'eronique Cayatte, Avishai Dekel, Peppo Gavazzi, Gopal Krishna,
and Gilles 
Theureau for useful discussions, and an anonymous referee for very
useful comments. We also thank
Fran\c{c}ois Bouchet,
Bruno Guiderdoni and coworkers for kindly providing us with their
$N$-body simulations, and Jeremy Blaizot for answering our
technical questions about the design and access to the simulations.
TS
acknowledges hospitality of the Institut d'Astrophysique de Paris where most of
this work was done, and she and GAM acknowledge Ewa {\L}okas for
hosting them at the CAMK in Warsaw, where part of this work was also done.
TS was supported by a fellowship
of the Ministerio de Educaci\'on, Cultura y Deporte of Spain.
We have used the {\sf HyperLEDA} database ({\tt http://leda.univ-lyon1.fr})
operated at the Observatoire de Lyon, France, {\sf SIMBAD}
({\tt http://simbad.u-strasbg.fr}), operated by the CDS in
Strasbourg, France, the
{\sf NASA/IPAC Extragalactic Database} ({\sf NED}, {\tt
http://nedwww.iap.caltech.edu}), 
which is operated by the Jet 
Propulsion Laboratory, California Institute of Technology, under contract
with the National Aeronautics and Space Administration, and
the {\sf GOLDMine} database operated by the Universita' degli Studi di
Milano-Bicocca.
\end{acknowledgements}

\appendix
\section{Median and half-width of the likelihood function for a set of
measurements with different errors}
\label{appsig}
For a single measurement of variable $X$ of value $x$ with uncertainty
$\sigma_i$, assuming  gaussian errors, 
the probability of measuring $x_i$ is
\begin{equation}
p_i(x_i|x) = {1\over \sqrt{2\pi} \sigma_i}\,\exp \left [-{(x_i-x)^2 \over
2\,\sigma_i^2} \right 
]
\ .
\label{pofi}
\end{equation}
The likelihood of measuring a set of $\{x_i\}$ given a true value $x$ is
\begin{eqnarray}
{\cal L} (\{x_i\}|x) &=& \prod_i p_i(x_i|x) \nonumber \\
&=&\prod_i \left ({1\over \sqrt{2\pi}
\sigma_i } \right ) \,\exp \left(-{S(x)\over2} \right ) \ ,
\label{likehood2}
\end{eqnarray}
with
\begin{equation}
S(x) = x^2\,\sum_i {1\over \sigma_i^2} - 2\, x\sum_i {x_i \over \sigma_i^2}
+ \sum_i {x_i^2 \over \sigma_i^2} \ .
\label{Sofx}
\end{equation}

{}From Bayes' theorem, the probability of having a true value $x$ is
\begin{equation}
P(x|\{x_i\}) = {f(x)\over g(\{x_i\})}\,{\cal L} \,(\{x_i\}|x) \ ,
\label{Pofx}
\end{equation}
where $f(x)$ is the a priori probability of having a true distance $x$ and 
$g(\{x_i\})$ is the a priori probability of measuring the set of $\{x_i\}$.
If the measurements are made with a selection function near unity, then $g
\simeq 1$. 

For simplicity, one can assume a uniform $f(x)$. 
The probability of the true $X$ being less than some value $x$ is then
\begin{eqnarray}
P(<x) &=& {\int_{-\infty}^x {\cal L} (\{x_i\}|x)\,\mathrm{d}x \over
\int_{-\infty}^\infty {\cal L} (\{x_i\}|x)\,\mathrm{d}x} \nonumber \\
&=& {
\int_{-\infty}^x \exp \left [-{x^2/ 2}\,\sum_i {1/ \sigma_i^2} +
x\,\sum_i {x_i / \sigma_i^2} \right ]\,\mathrm{d}x
\over
\int_{-\infty}^\infty \exp \left [-{x^2/ 2}\,\sum_i {1/ \sigma_i^2} +
x\,\sum_i {x_i / \sigma_i^2} \right ] \,\mathrm{d}x } \nonumber \\
&=& {1\over 2}\,\hbox{erfc} \left ({\nu\over \sqrt{2}} \right ) \ ,
\end{eqnarray}
where
\begin{equation}
\nu = {\sum_i x_i / \sigma_i^2 - x\,\sum_i 1/\sigma_i^2 \over \sqrt{\sum_i
1/\sigma_i^2}} \ .
\end{equation} 
Now $\hbox{erfc}(\nu/\sqrt{2}) / 2 = 0.16, 0.5, 0.84$ for $\nu = -1, 0, +1$,
so that the median $X$ is
\begin{equation}
x_{50} = {\sum_i x_i / \sigma_i^2 \over \sum_i 1/ \sigma_i^2} \ ,
\label{xmed}
\end{equation}
while the width equivalent to the standard deviation of a single gaussian is
\begin{equation}
\sigma = {1\over 2}\,\left (x_{84} - x_{16} \right )
= {1\over \sqrt{\sum_i 1/ \sigma_i^2}} \ .
\label{newsigma}
\end{equation}
Since ${\cal L}$ is an exponential function of $x$ 
(eq.~[\ref{likehood2}]), the mean and
mode will be equal to 
$x_{50}$.

Alternatively, one could
adopt an underlying physical model for $f(x)$. 
If $X$ is the distance modulus, for which
the gaussian p.d.f. of equation~(\ref{pofi}) should be adequate, one could use
a Virgocentric model, where the probability $f(x)$ stems from a density
distribution around Virgo, using for example an NFW model.
If the angle between the galaxy and Virgo, as seen by the observer, is
$\theta$, one has
\begin{eqnarray}
r^2 &=& D_{10}^2\,\left \{{\rm dex}(-0.4\,d) + {\rm dex}(-0.4\,d_V) \right.
\nonumber \\ 
&\mbox{}& \left. \qquad -
2\,\cos\theta \,{\rm dex}[-0.2\,(d+d_V)] \right \}
\ ,
\label{r2}
\end{eqnarray}
where $d$ and $d_V$ are the distance moduli to the galaxy and to Virgo,
respectively, and
$D_{10} = 10\,\rm pc$.
Hence, one has
\begin{eqnarray}
f(d) &\propto& \rho(r)\,{{\rm d}r\over {\rm d}d} \ ,\\
{{\rm d}r\over {\rm d}d} &=& {1\over 2\,r}\,{{\rm d}r^2\over {\rm d}d} 
\nonumber \\
&=& {0.2\,\ln 10\,D_{10}^2\over r}\,\nonumber \\
&\mbox{}& \times \left [\cos\theta\,{\rm dex}(-0.2(d+d_V)) - {\rm
dex}(-0.4d)\right] {\bf \,.}
\end{eqnarray}

The problem with this approach is that the underlying model,
i.e. $\rho(r)$, is uncertain:
Should one take a single NFW model at the location of M87? Should one make
$\rho(r)$ uniform at a radius where $\rho_\mathrm{NFW}(r)$ reaches the mean
mass density of the Universe?
Or better, should one replace this uniform background
with a collection of other halos, as expected in the hierarchical scenario?
Moreover, because the model requires a distance to M87, one would need to
iterate using a first guess on the distance to M87 (perhaps based upon the
estimate with a uniform $f(x)$).
These questions led us to assume in this paper a uniform distribution for
$f(x)$.


\begin{thebibliography}{54}
\expandafter\ifx\csname natexlab\endcsname\relax\def\natexlab#1{#1}\fi

\bibitem[{{Bettoni} {et~al.}(2003){Bettoni}, {Galletta}, \&
  {Garc{\'{\i}}a-Burillo}}]{BGG03}
{Bettoni}, D., {Galletta}, G., \& {Garc{\'{\i}}a-Burillo}, S. 2003, \aap, 405,
  5

\bibitem[{{Bevington} \& {Robinson}(1992)}]{Bevington92}
{Bevington}, P.~R. \& {Robinson}, D.~K. 1992, Data reduction and error analysis
  for the physical sciences (New York: McGraw-Hill, 2nd ed.)

\bibitem[{{Binggeli} {et~al.}(1985){Binggeli}, {Sandage}, \& {Tammann}}]{BST85}
{Binggeli}, B., {Sandage}, A., \& {Tammann}, G.~A. 1985, \aj, 90, 1681

\bibitem[{{Cardelli} {et~al.}(1989){Cardelli}, {Clayton}, \& {Mathis}}]{CCM89}
{Cardelli}, J.~A., {Clayton}, G.~C., \& {Mathis}, J.~S. 1989, \apj, 345, 245

\bibitem[{{Cayatte} {et~al.}(1994){Cayatte}, {Kotanyi}, {Balkowski}, \& {van
  Gorkom}}]{CKBvG94}
{Cayatte}, V., {Kotanyi}, C., {Balkowski}, C., \& {van Gorkom}, J.~H. 1994,
  \aj, 107, 1003

\bibitem[{{Cayatte} {et~al.}(1990){Cayatte}, {van Gorkom}, {Balkowski}, \&
  {Kotanyi}}]{CvGBK90}
{Cayatte}, V., {van Gorkom}, J.~H., {Balkowski}, C., \& {Kotanyi}, C. 1990,
  \aj, 100, 604

\bibitem[{{Chamaraux} {et~al.}(1980){Chamaraux}, {Balkowski}, \&
  {G\'erard}}]{CBG80}
{Chamaraux}, P., {Balkowski}, C., \& {G\'erard}, E. 1980, \aap, 83, 38

\bibitem[{{Cole} \& {Lacey}(1996)}]{CL96}
{Cole}, S. \& {Lacey}, C. 1996, \mnras, 281, 716

\bibitem[{{Cowley} {et~al.}(1982){Cowley}, {Crampton}, \& {McClure}}]{CCM82}
{Cowley}, A.~P., {Crampton}, D., \& {McClure}, R.~D. 1982, \apj, 263, 1

\bibitem[{{Davis} {et~al.}(1985){Davis}, {Efstathiou}, {Frenk}, \&
  {White}}]{DEFW85}
{Davis}, M., {Efstathiou}, G., {Frenk}, C.~S., \& {White}, S.~D.~M. 1985, \apj,
  292, 371

\bibitem[{{Ekholm} {et~al.}(2001){Ekholm}, {Baryshev}, {Teerikorpi}, {Hanski},
  \& {Paturel}}]{Ekholm+01}
{Ekholm}, T., {Baryshev}, Y., {Teerikorpi}, P., {Hanski}, M.~O., \& {Paturel},
  G. 2001, \aap, 368, L17

\bibitem[{{Ekholm} {et~al.}(2000){Ekholm}, {Lanoix}, {Teerikorpi}, {Fouqu{\'
  e}}, \& {Paturel}}]{ELTF00}
{Ekholm}, T., {Lanoix}, P., {Teerikorpi}, P., {Fouqu{\' e}}, P., \& {Paturel},
  G. 2000, \aap, 355, 835

\bibitem[{{Ferguson} \& {Binggeli}(1994)}]{FB94}
{Ferguson}, H.~C. \& {Binggeli}, B. 1994, \aapr, 6, 67

\bibitem[{{Gavazzi} {et~al.}(2003){Gavazzi}, {Boselli}, {Donati}, {Franzetti},
  \& {Scodeggio}}]{Gavazzi+03}
{Gavazzi}, G., {Boselli}, A., {Donati}, A., {Franzetti}, P., \& {Scodeggio}, M.
  2003, \aap, 400, 451

\bibitem[{{Gavazzi} {et~al.}(1999){Gavazzi}, {Boselli}, {Scodeggio}, {Pierini},
  \& {Belsole}}]{Gavazzi+99}
{Gavazzi}, G., {Boselli}, A., {Scodeggio}, M., {Pierini}, D., \& {Belsole}, E.
  1999, \mnras, 304, 595

\bibitem[{{Gavazzi} {et~al.}(2000){Gavazzi}, {Franzetti}, {Scodeggio},
  {Boselli}, \& {Pierini}}]{Gavazzi+00}
{Gavazzi}, G., {Franzetti}, P., {Scodeggio}, M., {Boselli}, A., \& {Pierini},
  D. 2000, \aap, 361, 863

\bibitem[{{Giovanelli} \& {Haynes}(1985)}]{GH85}
{Giovanelli}, R. \& {Haynes}, M.~P. 1985, \apj, 292, 404

\bibitem[{{Guiderdoni} \& {Rocca-Volmerange}(1985)}]{GRV85}
{Guiderdoni}, B. \& {Rocca-Volmerange}, B. 1985, \aap, 151, 108

\bibitem[{{Gunn} \& {Gott}(1972)}]{GG72}
{Gunn}, J.~E. \& {Gott}, J.~R. 1972, \apj, 176, 1

\bibitem[{{Hatton} {et~al.}(2003){Hatton}, {Devriendt}, {Ninin}, {Bouchet},
  {Guiderdoni}, \& {Vibert}}]{Hatton+03}
{Hatton}, S., {Devriendt}, J., {Ninin}, S., {et~al.} 2003, \mnras, 343, 75

\bibitem[{{Haynes} \& {Giovanelli}(1984)}]{HG84}
{Haynes}, M.~P. \& {Giovanelli}, R. 1984, \aj, 89, 758

\bibitem[{{Haynes} \& {Giovanelli}(1986)}]{HG86}
---. 1986, \apj, 306, 466

\bibitem[{{Jensen} {et~al.}(1998){Jensen}, {Tonry}, \& {Luppino}}]{JTL98}
{Jensen}, J.~B., {Tonry}, J.~L., \& {Luppino}, G.~A. 1998, \apj, 505, 111

\bibitem[{{Karachentsev} {et~al.}(2003){Karachentsev}, {Makarov}, {Sharina},
  {Dolphin}, {Grebel}, {Geisler}, {Guhathakurta}, {Hodge}, {Karachentseva},
  {Sarajedini}, \& {Seitzer}}]{Karachentsev+03}
{Karachentsev}, I.~D., {Makarov}, D.~I., {Sharina}, M.~E., {et~al.} 2003, \aap,
  398, 479

\bibitem[{{Keel}(1996)}]{Keel96}
{Keel}, W.~C. 1996, \pasp, 108, 917

\bibitem[{{Kenney} {et~al.}(1996){Kenney}, {Koopmann}, {Rubin}, \&
  {Young}}]{KKRY96}
{Kenney}, J.~D.~P., {Koopmann}, R.~A., {Rubin}, V.~C., \& {Young}, J.~S. 1996,
  \aj, 111, 152

\bibitem[{{Klypin} {et~al.}(2003){Klypin}, {Hoffman}, {Kravtsov}, \&
  {Gottl{\"o}ber}}]{KHKG03}
{Klypin}, A., {Hoffman}, Y., {Kravtsov}, A.~V., \& {Gottl{\"o}ber}, S. 2003,
  \apj, 596, 19

\bibitem[{{Kochanek} {et~al.}(2001){Kochanek}, {Pahre}, {Falco}, {Huchra},
  {Mader}, {Jarrett}, {Chester}, {Cutri}, \& {Schneider}}]{Kochanek+01}
{Kochanek}, C.~S., {Pahre}, M.~A., {Falco}, E.~E., {et~al.} 2001, \apj, 560,
  566

\bibitem[{{Koopmann} \& {Kenney}(1998)}]{KK98}
{Koopmann}, R.~A. \& {Kenney}, J.~D.~P. 1998, \apjl, 497, L75

\bibitem[{{Kotanyi} \& {Ekers}(1983)}]{KE83}
{Kotanyi}, C.~G. \& {Ekers}, R.~D. 1983, \aap, 122, 267

\bibitem[{{Kundu} \& {Whitmore}(2001)}]{KW01}
{Kundu}, A. \& {Whitmore}, B.~C. 2001, \aj, 121, 2950

\bibitem[{{Larson} {et~al.}(1980){Larson}, {Tinsley}, \& {Caldwell}}]{LTC80}
{Larson}, R.~B., {Tinsley}, B.~M., \& {Caldwell}, C.~N. 1980, \apj, 237, 692

\bibitem[{{Liu} \& {Kennicutt}(1995)}]{LK95}
{Liu}, C.~T. \& {Kennicutt}, R.~C. 1995, \apj, 450, 547

\bibitem[{{Mamon}(2000)}]{M00_IAP}
{Mamon}, G.~A. 2000, in 15th IAP Astrophys. Mtg., Dynamics of Galaxies: from
  the Early Universe to the Present, ed. F.~{Combes}, G.~A. {Mamon}, \&
  V.~{Charmandaris}, Vol. 197 (San Francisco: ASP), 377--388, astro-ph/9911333

\bibitem[{{Mamon} {et~al.}(2004){Mamon}, {Sanchis}, {Salvador-Sol{\' e}}, \&
  {Solanes}}]{MSSS03}
{Mamon}, G.~A., {Sanchis}, T., {Salvador-Sol{\' e}}, E., \& {Solanes}, J.~M.
  2004, \aap, in press, astro-ph/0310709

\bibitem[{{Mendes de Oliveira} {et~al.}(2003){Mendes de Oliveira}, {Amram},
  {Plana}, \& {Balkowski}}]{MAPB03}
{Mendes de Oliveira}, C., {Amram}, P., {Plana}, H., \& {Balkowski}, C. 2003,
  \aj, 126, in press

\bibitem[{{Naim} {et~al.}(1995){Naim}, {Lahav}, {Buta}, {Corwin}, {de
  Vaucouleurs}, {Dressler}, {Huchra}, {van den Bergh}, {Raychaudhury}, {Sodre},
  \& {Storrie-Lombardi}}]{Naim+95}
{Naim}, A., {Lahav}, O., {Buta}, R.~J., {et~al.} 1995, \mnras, 274, 1107

\bibitem[{{Neilsen} \& {Tsvetanov}(2000)}]{NT00}
{Neilsen}, E.~H. \& {Tsvetanov}, Z.~I. 2000, \apj, 536, 255

\bibitem[{{Ninin}(1999)}]{Ninin99}
{Ninin}, S. 1999, PhD thesis, Universit\'e de Paris XI

\bibitem[{{Rubin} {et~al.}(1991){Rubin}, {Hunter}, \& {Ford}}]{RHF91}
{Rubin}, V.~C., {Hunter}, D.~A., \& {Ford}, W.~Kent, J. 1991, \apjs, 76, 153

\bibitem[{{Rubin} {et~al.}(1999){Rubin}, {Waterman}, \& {Kenney}}]{RWK99}
{Rubin}, V.~C., {Waterman}, A.~H., \& {Kenney}, J.~D.~P. 1999, \aj, 118, 236

\bibitem[{{Sanchis} {et~al.}(2002){Sanchis}, {Solanes}, {Salvador-Sol{\' e}},
  {Fouqu{\' e}}, \& {Manrique}}]{Sanchis+02}
{Sanchis}, T., {Solanes}, J.~M., {Salvador-Sol{\' e}}, E., {Fouqu{\' e}}, P.,
  \& {Manrique}, A. 2002, \apj, 580, 164

\bibitem[{{Sandage} {et~al.}(1985){Sandage}, {Binggeli}, \& {Tammann}}]{SBT85}
{Sandage}, A., {Binggeli}, B., \& {Tammann}, G.~A. 1985, \aj, 90, 1759

\bibitem[{{Schlegel} {et~al.}(1998){Schlegel}, {Finkbeiner}, \&
  {Davis}}]{SFD98}
{Schlegel}, D.~J., {Finkbeiner}, D.~P., \& {Davis}, M. 1998, \apj, 500, 525

\bibitem[{{Solanes} {et~al.}(1996){Solanes}, {Giovanelli}, \& {Haynes}}]{SGH96}
{Solanes}, J.~M., {Giovanelli}, R., \& {Haynes}, M.~P. 1996, \apj, 461, 609

\bibitem[{{Solanes} {et~al.}(2001){Solanes}, {Manrique}, {Garc{\'{\i}}a-G{\'
  o}mez}, {Gonz{\' a}lez-Casado}, {Giovanelli}, \& {Haynes}}]{Solanes+01}
{Solanes}, J.~M., {Manrique}, A., {Garc{\'{\i}}a-G{\' o}mez}, C., {et~al.}
  2001, \apj, 548, 97

\bibitem[{{Solanes} {et~al.}(2002){Solanes}, {Sanchis}, {Salvador-Sol{\' e}},
  {Giovanelli}, \& {Haynes}}]{Solanes+02}
{Solanes}, J.~M., {Sanchis}, T., {Salvador-Sol{\' e}}, E., {Giovanelli}, R., \&
  {Haynes}, M.~P. 2002, \aj, 124, 2440

\bibitem[{{Spergel} {et~al.}(2003){Spergel}, {Verde}, {Peiris}, {Komatsu},
  {Nolta}, {Bennett}, {Halpern}, {Hinshaw}, {Jarosik}, {Kogut}, {Limon},
  {Meyer}, {Page}, {Tucker}, {Weiland}, {Wollack}, \& {Wright}}]{Spergel+03}
{Spergel}, D.~N., {Verde}, L., {Peiris}, H.~V., {et~al.} 2003, \apjs, 148, 175

\bibitem[{{Stauffer} {et~al.}(1986){Stauffer}, {Kenney}, \& {Young}}]{SKY86}
{Stauffer}, J.~R., {Kenney}, J.~D., \& {Young}, J.~S. 1986, \aj, 91, 1286

\bibitem[{{Teerikorpi} {et~al.}(1992){Teerikorpi}, {Bottinelli}, {Gouguenheim},
  \& {Paturel}}]{TBGP92}
{Teerikorpi}, P., {Bottinelli}, L., {Gouguenheim}, L., \& {Paturel}, G. 1992,
  \aap, 260, 17

\bibitem[{{Tonry} {et~al.}(2001){Tonry}, {Dressler}, {Blakeslee}, {Ajhar},
  {Fletcher}, {Luppino}, {Metzger}, \& {Moore}}]{Tonry+01}
{Tonry}, J.~L., {Dressler}, A., {Blakeslee}, J.~P., {et~al.} 2001, \apj, 546,
  681

\bibitem[{{Tully} \& {Fisher}(1977)}]{TF77}
{Tully}, R.~B. \& {Fisher}, J.~R. 1977, \aap, 54, 661

\bibitem[{{Tully} \& {Shaya}(1984)}]{TS84}
{Tully}, R.~B. \& {Shaya}, E.~J. 1984, \apj, 281, 31

\bibitem[{{Young} \& {Currie}(1995)}]{YC95}
{Young}, C.~K. \& {Currie}, M.~J. 1995, \mnras, 273, 1141

\end{thebibliography}

\end{document}